\documentclass{article}

\usepackage{natbib}

 \usepackage[accepted,nohyperref]{icml2013}

\usepackage{graphicx}
\usepackage{xcolor}

\usepackage[
protrusion=true, 
expansion=true,
tracking=true,
kerning=true,
spacing=true,
letterspace=50,
shrink=40, 
factor=1000
]{microtype} 

\usepackage{comment}
\colorlet{TufteRed}{red!80!black}

\definecolor{theblue}{RGB}{0,0,180}
\colorlet{thered}{TufteRed}

\usepackage{etoolbox}
\let\bbordermatrix\bordermatrix
\patchcmd{\bbordermatrix}{8.75}{4.75}{}{}
\patchcmd{\bbordermatrix}{\left(}{\left[}{}{}
\patchcmd{\bbordermatrix}{\right)}{\right]}{}{}
\usepackage{longtable}
\usepackage{booktabs}

\usepackage[notheorems]{rr-math}
\newcommand\TTT{\rule{0pt}{3.2ex}}
\newcommand\BBB{\rule[-1.4ex]{0pt}{0pt}}

\newcommand\TT{\rule{0pt}{2.3ex}}
\newcommand\BB{\rule[-1.0ex]{0pt}{0pt}}

\usepackage{graphicx}
\usepackage{amssymb}
\usepackage{amsmath}
\usepackage{multirow}
\usepackage{epstopdf}
\usepackage{tabularx}
\usepackage{booktabs}
\usepackage{paralist}
\usepackage{url}
\usepackage{graphicx}
\usepackage{xspace}
\usepackage{listings}

\usepackage{graphicx}
\usepackage{balance}  

\usepackage{amssymb}
\usepackage{amsmath}
\usepackage{amsfonts}

\usepackage{algorithm}
\usepackage{algorithmicx}
\usepackage{algpseudocode}

\usepackage{listings}

\newtheorem{mydef}{Definition}
\newtheorem{myproblem}{Problem}

\newcommand{\eol}{\end{enumerate}\setlength{\itemsep}{-\parsep}}

\providecommand{\eg}{\emph{e.g.}\xspace}
\providecommand{\ie}{\emph{i.e.}\xspace}
\providecommand{\etal}{\emph{et al.}\xspace}

\providecommand{\wrt}{\emph{w.r.t.}\xspace}

\relax

\def\E{{\mathbb E}}

\usepackage{wrapfig}
\usepackage{xspace}
\usepackage{csquotes}

\newlength{\commentWidth}
\newcommand{\bspacing}{\begin{spacing}{1.1}}
\newcommand{\espacing}{\end{spacing}}

\definecolor{plotblue}{RGB}	{30,144,255}
\definecolor{plotgreen}{RGB}	{50,205,50}
\definecolor{plotred}{RGB}	{220,20,60}

\definecolor{myyellow}{RGB}{255,255,204}
\definecolor{myred}{RGB}{255,204,204}
\definecolor{myblue}{RGB}{0,200,255}
\definecolor{mygreen}{RGB}{80,220,80}

\colorlet{TufteRed}{red!80!black}
\definecolor{theblue}{RGB}{0,0,180}
\colorlet{thered}{TufteRed}

\definecolor{gray}{RGB}{150,150,150}
\usepackage{textcase}
\usepackage{soul}

\newcommand{\be}{\begin{equation}}
\newcommand{\ee}{\end{equation}}
\newcommand{\bea}{\begin{eqnarray}}
\newcommand{\eea}{\end{eqnarray}}
\providecommand{\bEqArray}{\begin{eqnarray}\begin{array}{l}}
\providecommand{\eEqArray}{\end{array}\end{eqnarray}}

\newcommand{\bit}{\begin{itemize}}
\newcommand{\eit}{\end{itemize}}

\definecolor{lightgray}{rgb}{0.93,0.93,0.93}

\definecolor{lightblue}{rgb}{0.5,0.90,1.0}
\definecolor{lightgreen}{rgb}{0.5,0.92,0.5}
\definecolor{lightred}{rgb}{0.98,0.5,0.5}
\definecolor{lightyellow}{rgb}{1,0.90,0.40}

\newcommand{\pgd}{\ensuremath{\textsc{pgd}}}
\newcommand{\orca}{\ensuremath{\textsc{orca}}}
\newcommand{\rage}{\ensuremath{\textsc{rage}}}
\newcommand{\fanmod}{\ensuremath{\textsc{fanmod}}}

\usepackage{comment}
\graphicspath{{graphics/}{../graphics/}{./}}

\usepackage{array}
\newcolumntype{P}[1]{>{\centering\arraybackslash}p{#1}}
\newcolumntype{M}[1]{>{\centering\arraybackslash}m{#1}}

\usepackage{hyperref}
  \definecolor{mydarkblue}{rgb}{0,0.08,0.45}
  \hypersetup{ %
    colorlinks=true,
    linkcolor=mydarkblue,
    citecolor=mydarkblue,
    urlcolor=mydarkblue,
bookmarks=false}
\usepackage{textcase}
\usepackage{multirow}
\definecolor{myyellow}{RGB}{255,255,204}
\definecolor{myred}{RGB}{255,204,204}
\definecolor{myblue}{RGB}{0,200,255}
\definecolor{mygreen}{RGB}{80,220,80}

\algblockdefx[parallel]{parallelfor}{parallelend}
[1][]{\textbf{parallel for} #1}
{\textbf{end parallel}}

\newcommand{\ds}\displaystyle
\newcommand{\mbb}\mathbb
\newcommand{\mc}\mathcal
\newcommand{\del}\nabla

\newcommand{\beqstar}{\begin{eqnarray*}}
\newcommand{\eeqstar}{\end{eqnarray*}}

\usepackage{amssymb}
\usepackage{amsmath}
\usepackage{amsfonts}
\newcommand{\eat}[1]{}

\usepackage{nicefrac}

\algnotext{EndFor}
\algnotext{EndIf}
\algnotext{EndProcedure}
\algnotext{EndParallel}

\graphicspath{{./}{./figures/}}

\usepackage{mathdots}
\usepackage{setspace}

\graphicspath{{./}{./images/}}
\newcolumntype{H}{>{\setbox0=\hbox\bgroup}c<{\egroup}@{}}

\algblockdefx[parallel]{parfor}{endpar}[1][]{$\textbf{parallel for}$ #1 $\textbf{do}$}{$\textbf{end parallel}$}
\usepackage{amssymb}
\definecolor{gray}{RGB}{20,20,20}
\definecolor{greencm}{RGB}{0,153,0}

\colorlet{TufteRed}{red!80!black}
\colorlet{thered}{TufteRed}

\newcommand{\KK}{\mathbb{K}}

\renewcommand{\Std}{\mathbb{S}}

\newcommand{\G}{\ensuremath{{\mathcal{H}}}}
\newcommand{\g}{\ensuremath{{H}}}

\newcommand{\mychoose}[2]{\left( \begin{smallmatrix} #1 \\ #2 \end{smallmatrix} \right)} 

\providecommand{\X}{\ensuremath{\mX}}

\renewcommand{\E}{\ensuremath{E}}
\newcommand{\e}{\ensuremath{e}}

\newcommand{\C}{\ensuremath{C}}

\providecommand{\p}{\ensuremath{p}}

\newcommand{\dmax}{\ensuremath{\Delta}} 
\renewcommand{\th}{\ensuremath{\rm \mathit{th}}}

\renewcommand{\p}{\ensuremath{p}} 
 
\newcommand{\hash}{\ensuremath{{\boldsymbol\Psi}}}

\renewcommand{\S}{\ensuremath{{S}}}

\newcommand{\tri}{\ensuremath{{T}}}
\renewcommand{\O}{\ensuremath{\mathcal{O}}}
\renewcommand{\bigO}[1]{\ensuremath{\mathcal{O}(#1)}}

\newcommand{\data}[2]{{\mathsf{#1}\text{--}{\mathbf{\tt #2}}}}

\newcommand{\m}{\ensuremath{M}}
\newcommand{\n}{\ensuremath{N}}

\renewcommand{\d}{\ensuremath{d}}

\newcommand{\N}{\ensuremath{\Gamma}} 
\newcommand{\Ng}{\ensuremath{\boldsymbol \Gamma}}

\newcommand{\I}{\ensuremath{I}}
\newcommand{\ngpu}{\ensuremath{p}}

\renewcommand{\b}{\ensuremath{b}}

\newenvironment{myitemize}{%
  \begin{compactenum}[\,\,{\small $\mathbf{\boldsymbol-}$}\,\,\,]
  \smallskip
  \parskip=6pt
  
}{%
  \medskip
  \end{compactenum}
}

\renewenvironment{itemize}{%
  \begin{compactenum}[\,\,{\small $\bullet$}\,]
  \smallskip
  \parskip=6pt
  
}{%
  \medskip
  \end{compactenum}
}

\icmltitlerunning{Hybrid CPU-GPU Framework for Network Motifs}
\begin{document} 

\twocolumn[

\icmltitle{Hybrid CPU-GPU Framework for Network Motifs}

\icmlauthor{Ryan A. Rossi}{rrossi@parc.com}
\icmlauthor{Rong Zhou}{rzhou@parc.com}
\icmladdress{Palo Alto Research Center (PARC),  3333 Coyote Hill Rd, Palo Alto, CA 94304 USA}

%


\icmlkeywords{
Graphlets;  
induced subgraphs;
network motifs; 
graphlet decomposition; 
GPU computing;
multi-GPU; 
heterogeneous computing; 
parallel algorithms; 
graph mining; 
role discovery; 
relational learning; 
graph classification;
graph kernels;
feature learning;
orbits;
node and edge embedding;
graph representation learning
}

\vskip 0.3in
]

\begin{abstract}
Massively parallel architectures such as the GPU are becoming increasingly important due to the recent proliferation of data. In this paper, we propose a key class of hybrid parallel graphlet algorithms that leverages multiple CPUs and GPUs simultaneously for computing k-vertex induced subgraph statistics (called graphlets). In addition to the hybrid multi-core CPU-GPU framework, we also investigate single GPU methods (using multiple cores) and multi-GPU methods that leverage all available GPUs simultaneously for computing induced subgraph statistics. Both methods leverage GPU devices only, whereas the hybrid multi-core CPU-GPU framework leverages all available multi-core CPUs and multiple GPUs for computing graphlets in large networks. Compared to recent approaches, our methods are orders of magnitude faster, while also more cost effective enjoying superior performance per capita and per watt. In particular, the methods are up to 300 times faster than a recent state-of-the-art method. To the best of our knowledge, this is the first work to leverage multiple CPUs and GPUs simultaneously for computing induced subgraph statistics.
\end{abstract}

\renewcommand{\d}{\ensuremath{d}}

\section{Introduction}\label{sec:intro}
\noindent
Graphlets are small $k$-vertex induced subgraphs 
and are important for many predictive and descriptive modeling tasks~\cite{prvzulj2004modeling,milenkoviae2008uncovering,hayes2013graphlet} in a variety of disciplines including 
bioinformatics~\cite{vishwanathan2010graph,shervashidze2009efficient},
cheminformatics~\cite{rupp2010graph}, and image processing and computer vision~\cite{zhang-image-categorization-via-graphlets,zhang2013probabilistic}.
Given a network $G$, our approach counts the frequency of each $k$-vertex induced subgraph patterns (See Table~\ref{table:graphlet-notation}).
These counts represent powerful features that succinctly characterize the fundamental network structure~\cite{shervashidze2009efficient}. 
Indeed, it has been shown that such features accuratly capture the local network structure in a variety of domains~\cite{Holland_Lein,faust2010puzzle,frank1988triad}. 
As opposed to global graph parameters such as diameter for which two or more networks may have global graph parameters that are nearly identical, yet their local structural properties may be significantly different.

Despite the practical importance of graphlets, existing algorithms are slow and are limited to small networks (e.g., even modest graphs can take days to finish), require vast amounts of memory/space-inefficient, are inherently sequential (inefficient and difficult to parallelize), and have many other issues.
Overcoming the slow performance and scalability limitations of existing methods is perhaps the most important and challenging problem remaining. 
This work proposes a fast hybrid parallel algorithm for computing both \emph{connected} and \emph{disconnected} subgraph statistics (of size $k$) including macro-level statistics for the global graph $G$ as well as micro-level statistics for individual graph elements such as edges.
Furthermore, a number of important machine learning tasks are likely to benefit from the proposed methods, including graph anomaly detection~\cite{noble2003graph,akoglu2014graph}, entity resolution~\cite{bhattacharya2006entity}, as well as features for improving community detection~\cite{schaeffer2007graph}, role discovery~\cite{rossi2014roles}, and relational classification~\cite{getoor2007introduction}.

The recent rise of Big Data has brought many challenges and opportunities~\cite{boyd2012critical,ahmed2014graph,ahmed12socialnets}.
Recent heterogeneous high performance computing (HPC) architectures offer viable platforms for addressing the computational challenges of mining and learning with big graph data. 
General-purpose graphics processing units (GPGPUs)~\cite{brodtkorb2013graphics} are becoming increasingly important with applications in scientific computing, machine learning, and many others~\cite{gharaibeh2013energy}.
Heterogeneous (hybrid) computing architectures consisting of \emph{multi-core CPUs and multiple GPUs} are an attractive alternative to traditional homogeneous HPC clusters (HPCC). 
Despite the impressive theoretical performance achievable by such hybrid architectures, 
leveraging this computing power remains an extremely challenging problem.

Graphics processing units (GPUs) offer cost-effective high-performance solutions for computation-intensive data mining applications. 
As an example, the Nvidia Titan Black GPU has 2880 cores capable of performing over 5 trillion floating-point operations per second (TFLOPS).
For comparison a Xeon E5-2699v3 processor can perform about 0.75 TFLOPS, but can cost 4x as much. 
Besides TFLOPS, GPUs also enjoy a significant advantage over CPUs in terms of memory bandwidth, which is more important for data-intensive algorithms such as graphlet decomposition. 
For Titan Black, its maximum memory bandwidth is 336 GB/sec; whereas E5-2699v3's is only 68 GB/sec. 
Higher memory bandwidth means more graph edges can be traversed on the GPU than the CPU in the same amount of time, which is one of the main reasons why our approach can outperform the state-of-the-art.

\begin{figure}[t!]
\centering
\includegraphics[width=0.9\linewidth]{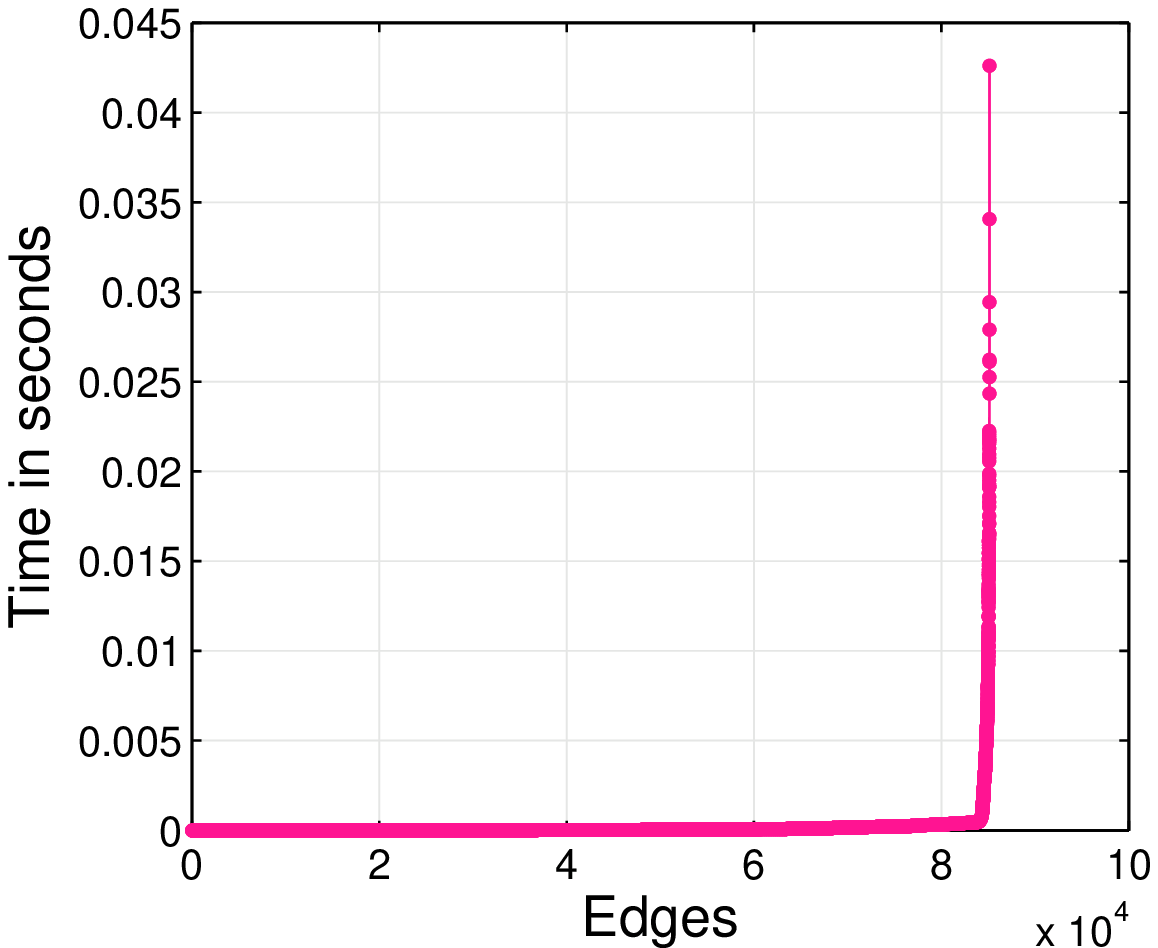}
\caption{The distribution of graphlet computation times for the edge neighborhoods obey a power-law.
The time taken to count $k=\{2,3,4\}$ graphlets for each edge is shown above and clearly obeys a power-law distribution ($\data{tech}{internet\text{-}as}$). 
}
\label{fig:graphlet-edge-powerlaw}
\end{figure}

However, unlike the \emph{single GPU-based approach} proposed in~\cite{milinkovic14contribution}, 
we adopt a hybrid parallelization strategy that exploits the complementary features of \emph{multiple GPUs and CPUs} to achieve significantly better performance.
We begin with the key observation that the amount of work required to compute graphlets for each edge in $G$ obeys a power-law (See Figure~\ref{fig:graphlet-edge-powerlaw}).
Strikingly, a handful of edges require a lot of work to determine the local graphlet counts (due to the density and structure of the local edge neighborhood), whereas the vast majority of other edges require only a small amount of work.
Such heterogeneity can cause significant load balancing issues, especially for GPUs that are designed to solve problems with uniform workloads (e.g., dense matrix-matrix multiplications). 
This motivates our hybrid approach that dynamically divides up the work between GPU and CPU, to reduce inter-processor communication, synchronization and data transfer overheads.

This work demonstrates that parallel graphlet methods designed for heterogeneous computing architectures consisting of many multi-core CPUs and multiple GPUs can significantly improve performance over GPU-only and CPU-only approaches.
In particular, our hybrid CPU-GPU framework is designed to leverage the advantages and key features offered by each type of processing unit (multi-core CPUs and GPUs).
As such, our approach leverages the fact that graphlets can be computed via $\m$ independent edge-centric neighborhood computations.
Therefore, the method dynamically distributes the edge-centric graphlet computations to either a CPU or a GPU.
In particular, the edge-centric graphlet computations that are fundamentally unbalanced and highly-skewed are given to the CPUs whereas the GPUs work on the more well-balanced and regular edge neighborhoods (See Figure~\ref{fig:graphlet-cpu-gpu-runtime}).
These approaches capitalize on the fact that GPUs are generally best for computations that are well-balanced and regular, whereas CPUs are designed for a wide variety of applications and thus more flexible~\cite{lee2010debunking,malony2011parallel}.
Our approach also leverages \emph{dynamic load balancing} and \emph{work stealing strategies} to ensure all GPU and CPU workers (cores) remain fully utilized.

\section{Related Work} \label{sec:related-work}
\noindent
Recently, Milinkovi{\'c}~\etal~\cite{milinkovic14contribution} proposed a GPU algorithm for counting graphlets based on a recent sequential graphlet algorithm called $\orca$~\cite{orca}.
However, this paper is fundamentally different.
First and foremost, that approach is not hybrid and is only able to use a single GPU for computing graphlets.
In addition, that work focuses on computing connected graphlets only, whereas we compute both connected and disconnected induced subgraphs.
Moreover, that approach computes graphlets for each vertex in parallel (vertex-centric), whereas our methods are naturally edge-centric and search edge neighborhoods in parallel.
Furthermore, that work does not provide any comparison to understand the utility and speedup (if any) offered by their approach.

In this work, we propose a heterogeneous graphlet framework for hybrid multi-GPU and CPU systems that leverages all available GPUs and CPUs for efficient graphlet counting.
Our single-GPU, multi-GPU, and hybrid CPU-GPU algorithms are largely inspired by the recent \emph{state-of-the-art} parallel (CPU-based) exact graphlet decomposition algorithm called $\pgd$~\cite{pgd-kais,nkahmed:icdm15}, 
which is known to be significantly faster and more efficient than other methods including $\rage$~\cite{rage}, $\fanmod$~\cite{fanmod}, and $\orca$~\cite{orca}.
Moreover, $\pgd$ has been parallelized for multi-core CPUs and is publicly available\footnote{\href{http://github.com/nkahmed/pgd}{$\mathtt{www.github.com/nkahmed/pgd}$}}.
Our approach is evaluated against $\pgd$ and a recent $\orca$-GPU approach in Section~\ref{sec:exp}.

{
\def\tabularxcolumn#1{m{#1}}
\providecommand{\graphletSZ}{0.045} 
\setlength{\tabcolsep}{3.0pt}
\begin{table}[h!]
\caption{Summary of the graphlet notation}
\centering
\small
\smallskip
\label{table:graphlet-notation}
\begin{tabularx}{1.0\linewidth}{@{}l c H c c l c c l HH HHH H H HHH HH @{}}
\midrule

\multicolumn{9}{p{0.99\linewidth}}
{\small
Graphlets are grouped by number of vertices ($k$-graphlets) and categorized into connected and disconnected graphlets.
Connected graphlets of each size are then ordered by density.
The complement of each connected graphlet is shown on the right and represent the disconnected graphlets.
Note $4$-path is a self-complementary$^{\star}$.
Graphlets of size $k$=2 are included for completeness.
\vspace{1mm}}
\\

\toprule
& $\mathbf{k}$ & &
\multicolumn{3}{l}{\textsc{Connected}} &
\multicolumn{3}{l}{\textsc{Disconnected}} &
\\
\midrule 

\renewcommand{\arraystretch}{1.4}
\TTT\BBB
\multirow{12}{*}{\rotatebox{90}{\mbox{
\textsc{\sc graphlets}
}}}
&
\multirow{1}{*}{\rotatebox{0}{\mbox{
$\mathbf{2}$
}}}
& 
&  \includegraphics[scale=\graphletSZ]{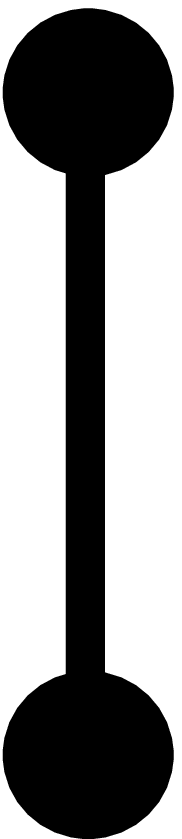} 
& $\g_1$  & edge 
&  \includegraphics[scale=\graphletSZ]{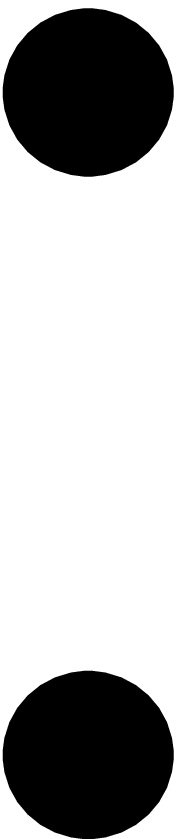} 
& $\g_2$  & 2-node-indep.
\\ 

\cmidrule{2-9}
&
\TTT\BBB
\multirow{2}{*}{\rotatebox{0}{\mbox{
$\mathbf{3}$ 
}}} 
& 
&  \includegraphics[scale=\graphletSZ]{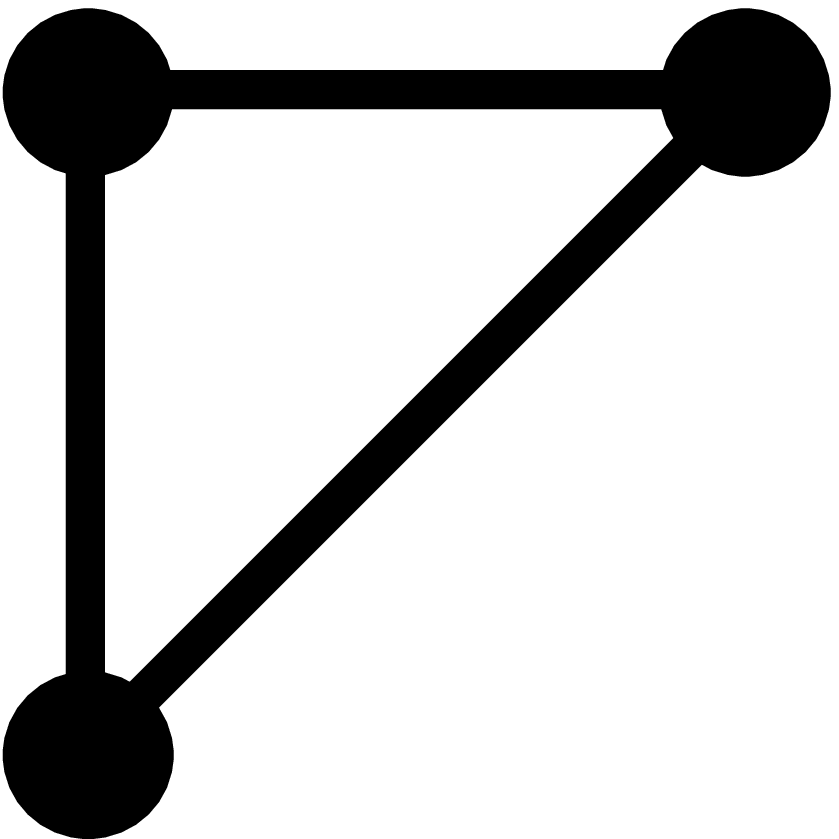} 
& $\g_3$  & triangle 
&  \includegraphics[scale=\graphletSZ]{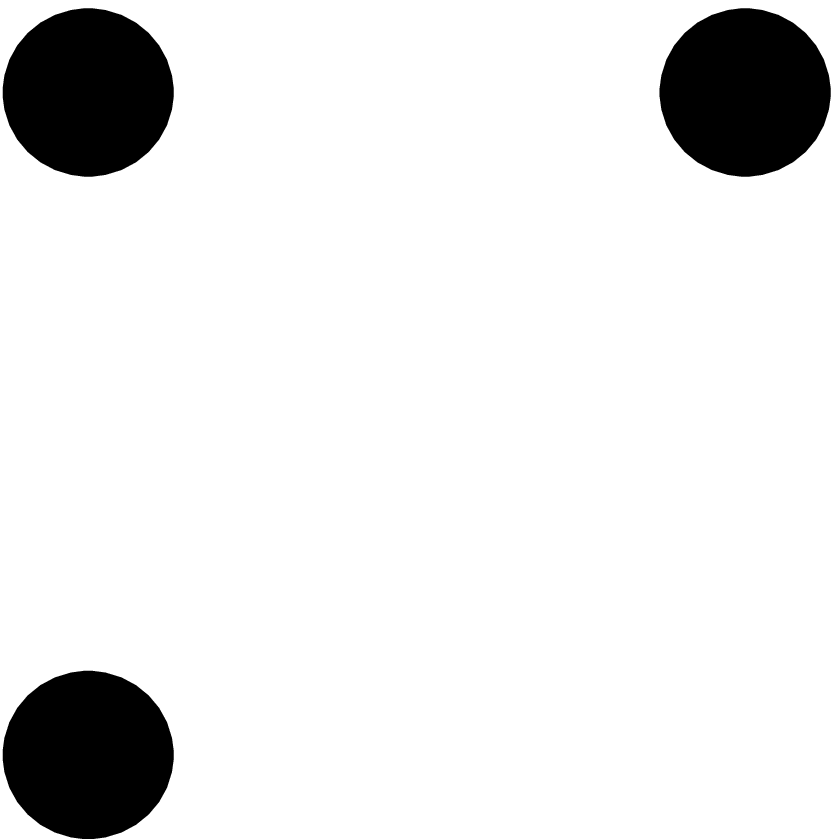} 
& $\g_6$  & 3-node-indep.
\\ 

\TTT\BBB
& &
&  \includegraphics[scale=\graphletSZ]{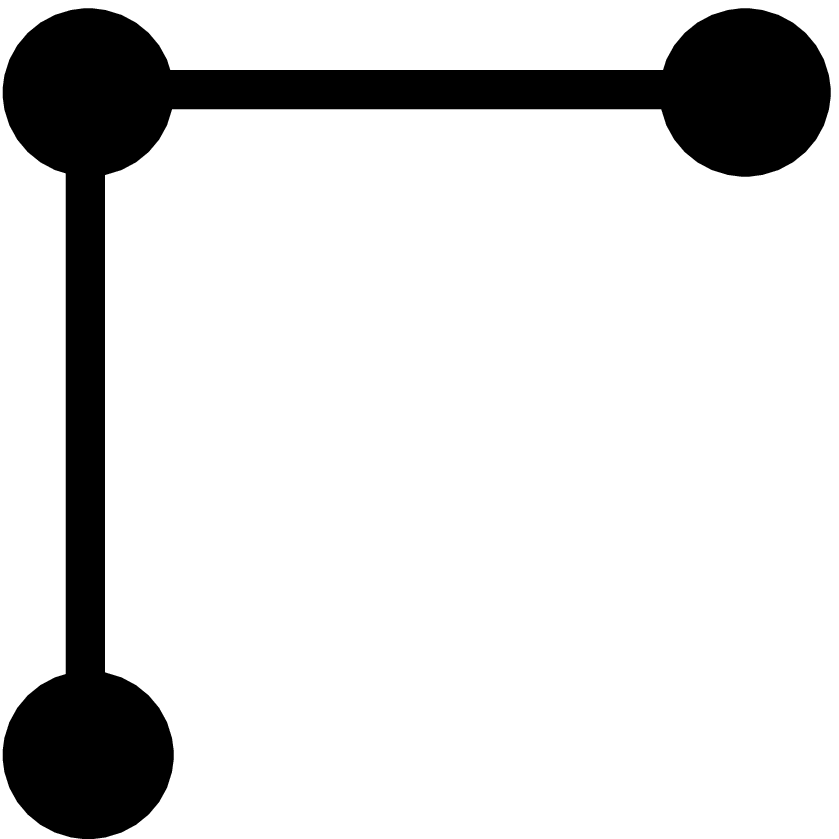} 
& $\g_4$  & 2-star 
&  \includegraphics[scale=\graphletSZ]{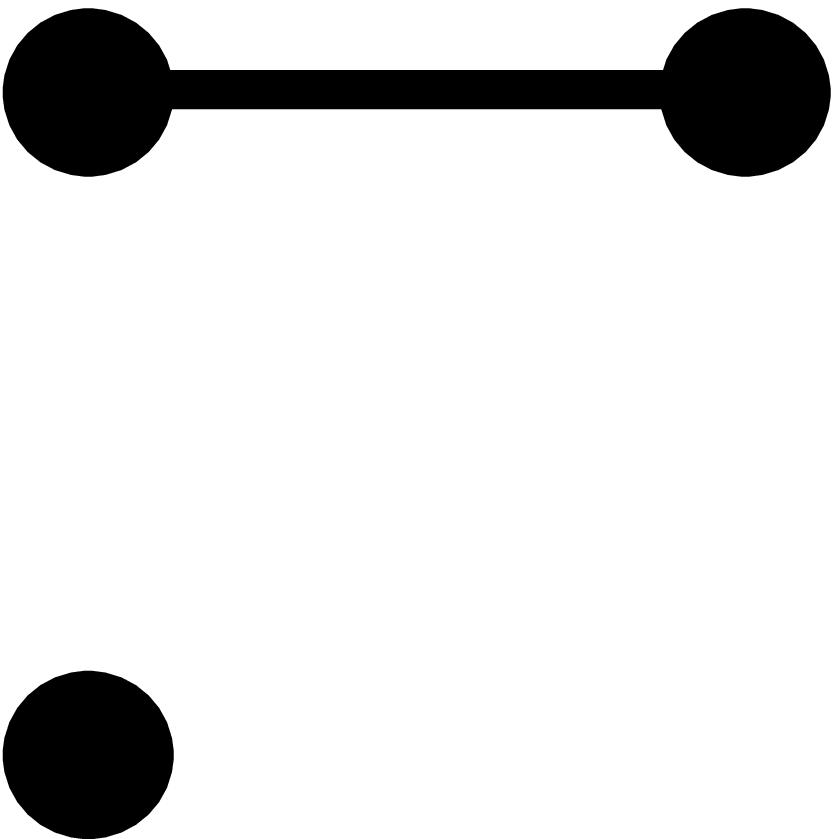} 
& $\g_5$  & 3-node-1-edge 
\\ 

\cmidrule{2-9}
\TTT\BBB
&
\multirow{7}{*}{\mbox{
$\mathbf{4}$
}}
& &  \includegraphics[scale=\graphletSZ]{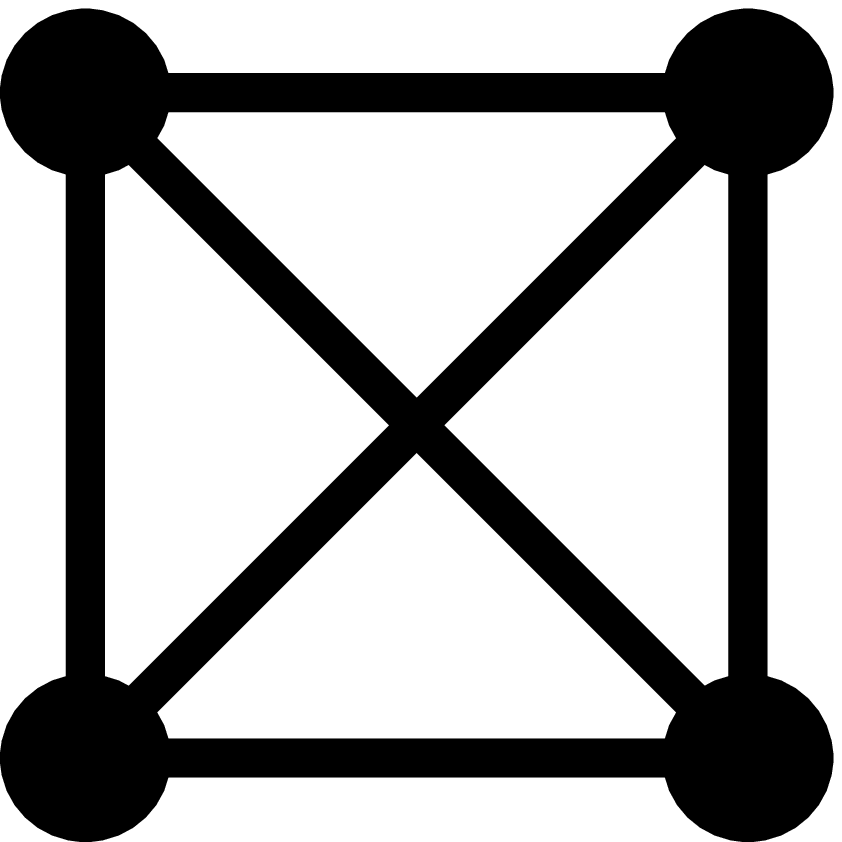} 
& $\g_7$  & 4-clique 
&  \includegraphics[scale=\graphletSZ]{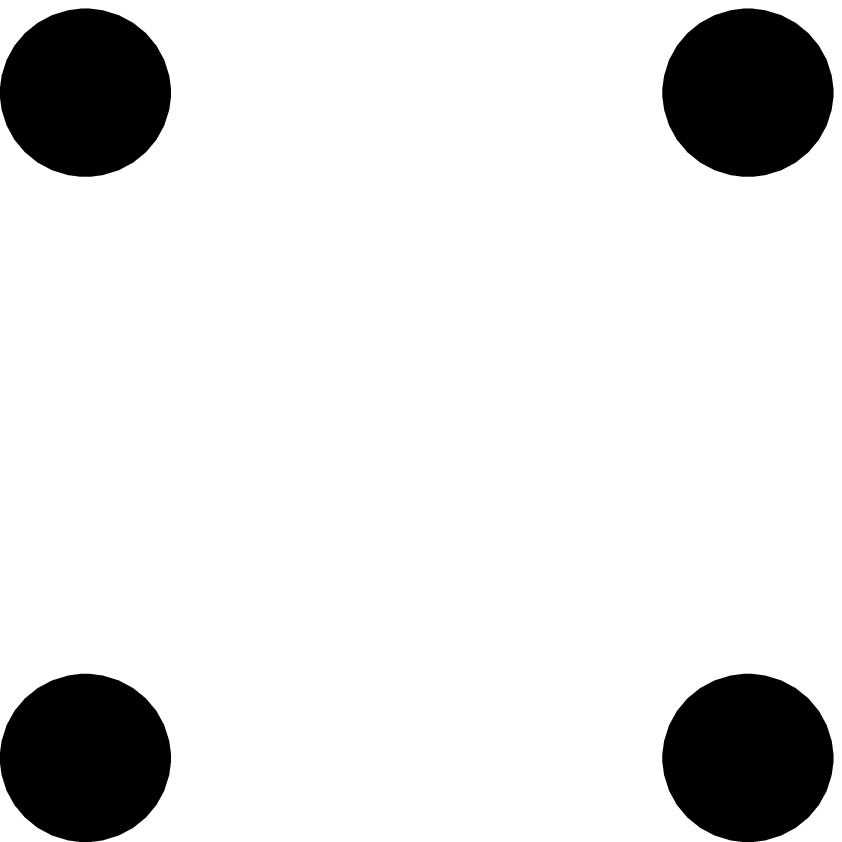} 
& $\g_{17}$  & 4-node-indep.
\\ 

\TTT\BBB
& &
 &  \includegraphics[scale=\graphletSZ]{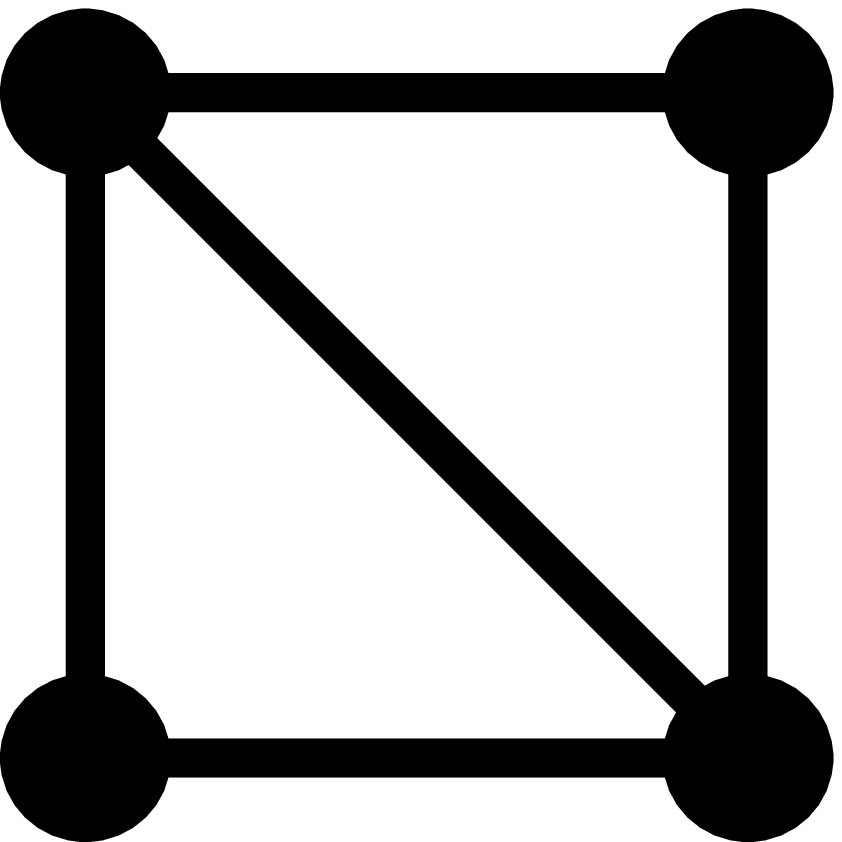} 
& $\g_8$  & 
chordal-cycle
&  \includegraphics[scale=\graphletSZ]{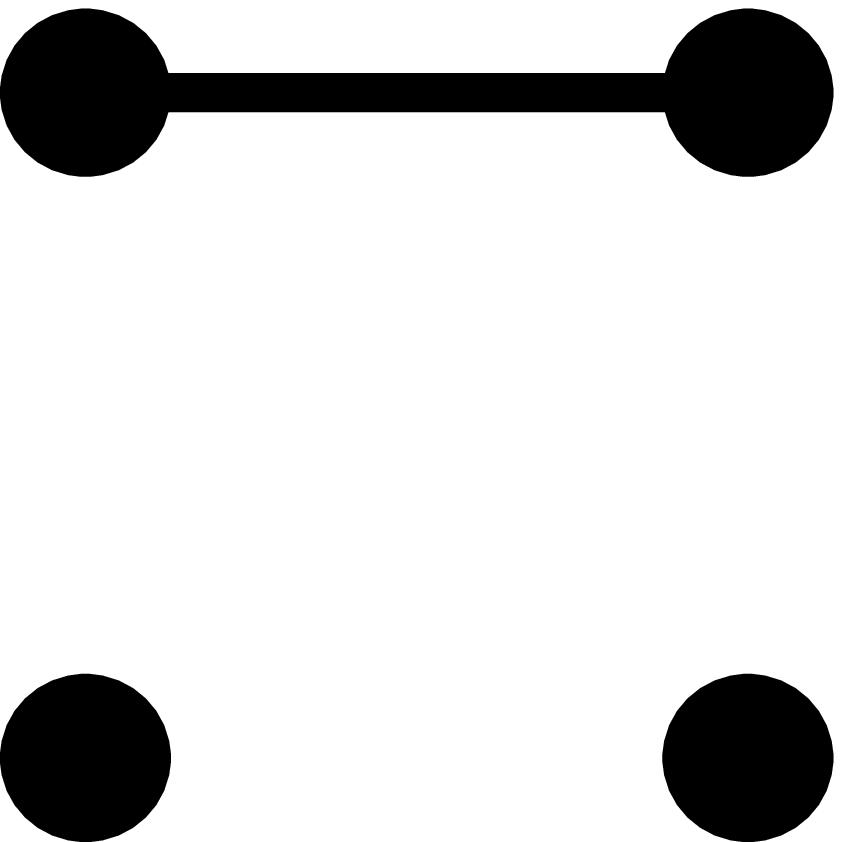} 
& $\g_{16}$  & 4-node-1-edge
\\ 

\TTT\BBB
& &
&  \includegraphics[scale=\graphletSZ]{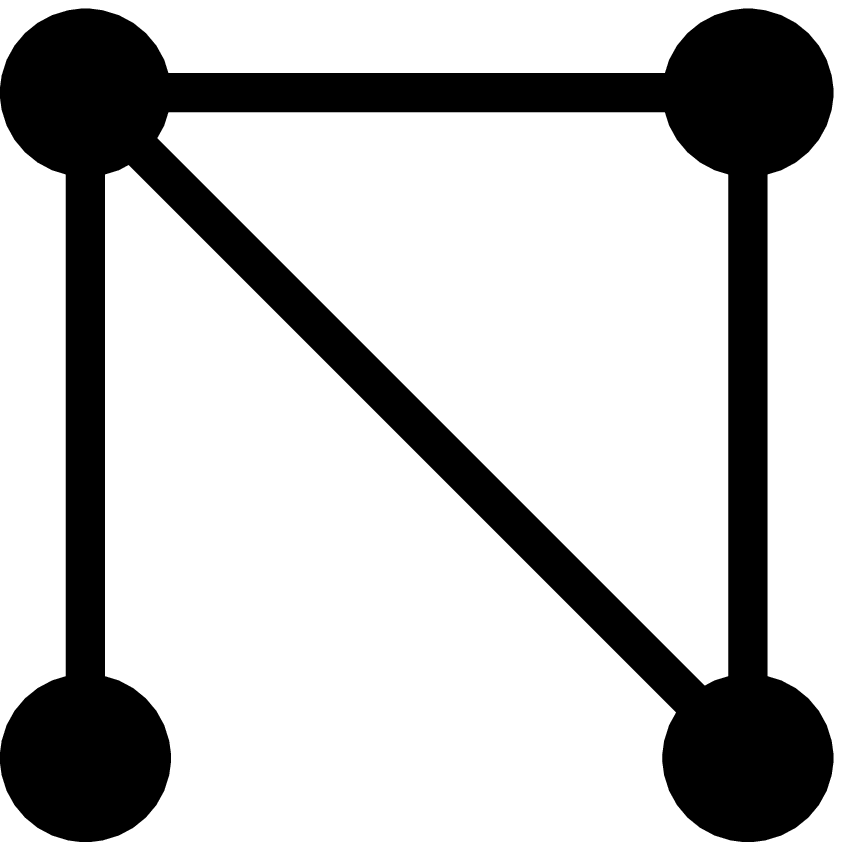} 
& $\g_9$  & 
tailed-triangle
&  \includegraphics[scale=\graphletSZ]{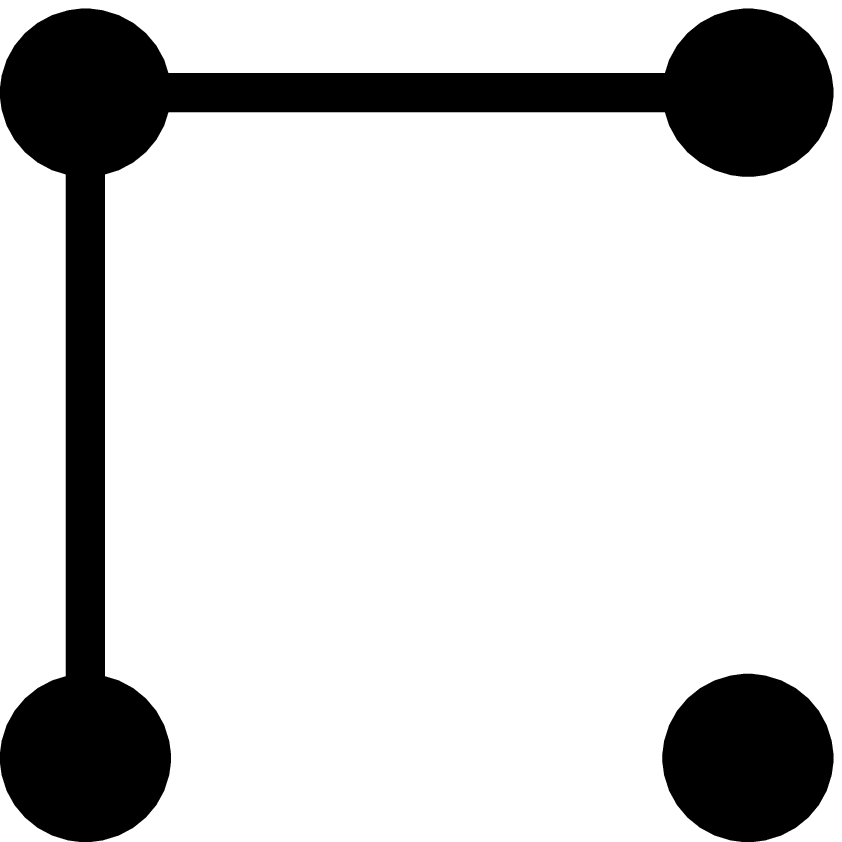} 
& $\g_{15}$  & 4-node-2-star
\\

\TTT\BBB
& &
&  \includegraphics[scale=\graphletSZ]{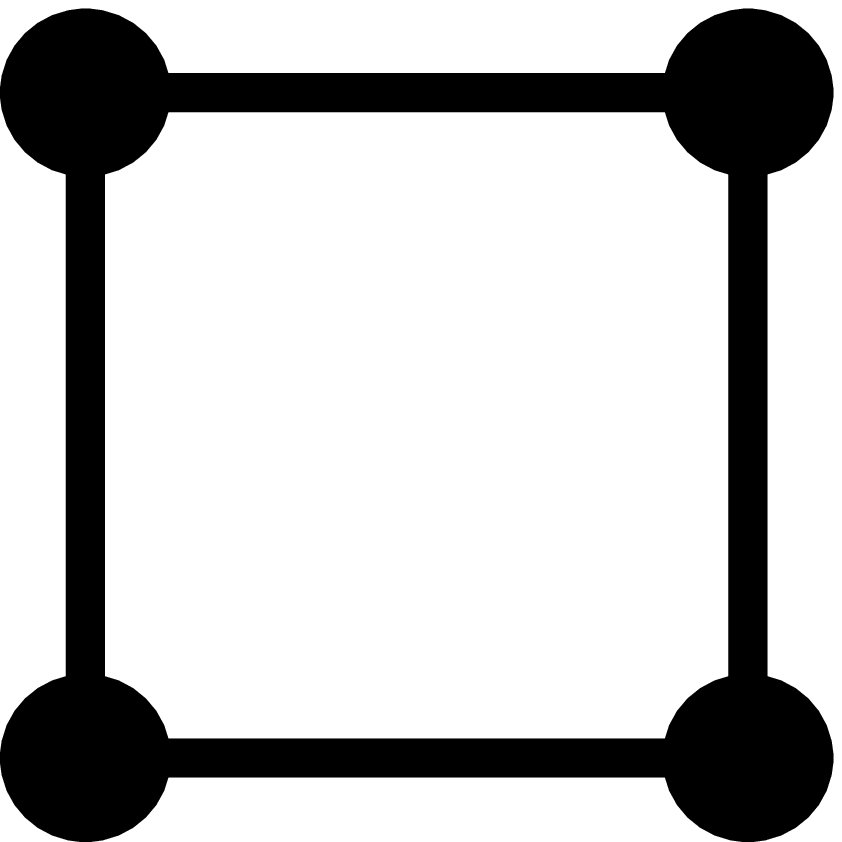} 
& $\g_{10}$  & 4-cycle 
&  \includegraphics[scale=\graphletSZ]{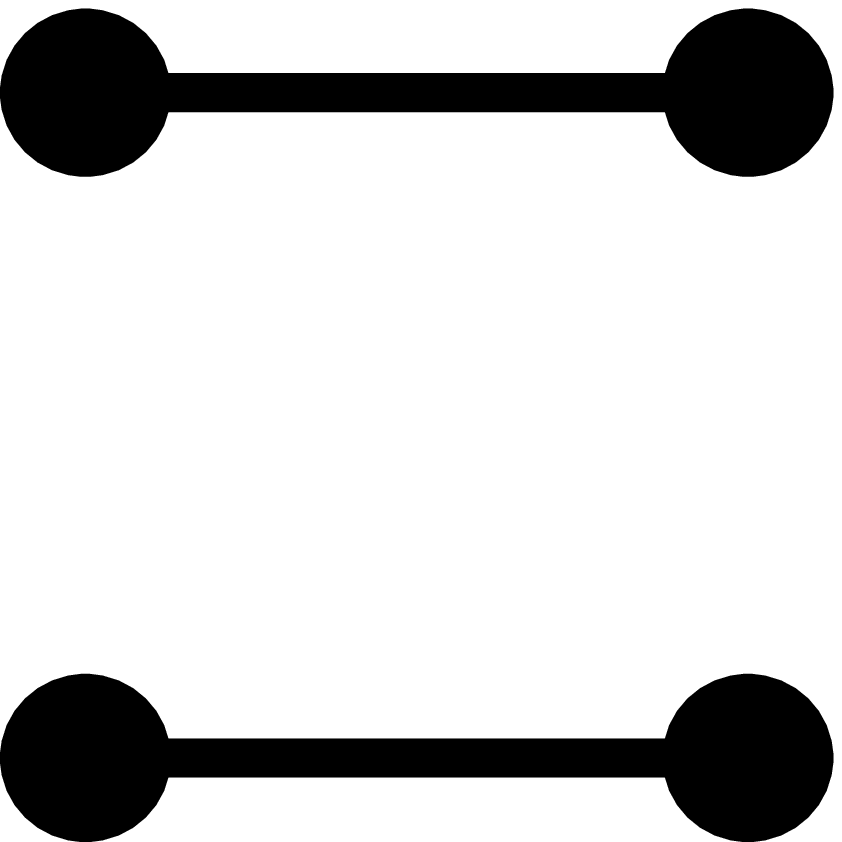} 
& $\g_{14}$  & 4-node-2-edge 
\\

\TTT\BBB
& &
&  \includegraphics[scale=\graphletSZ]{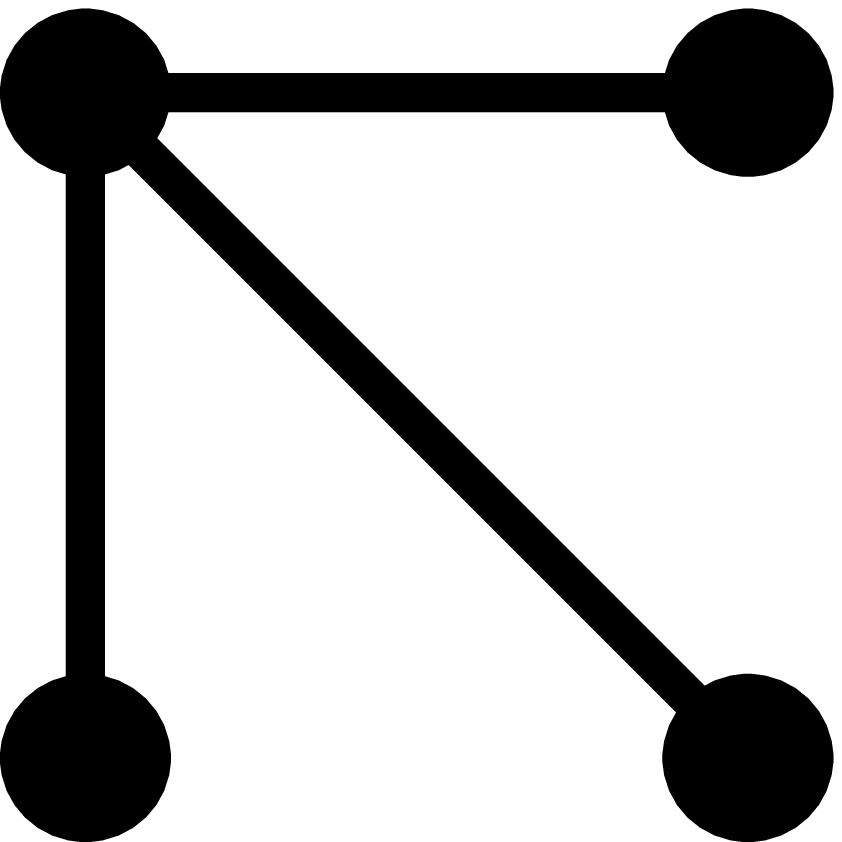} 
& $\g_{11}$  & 3-star 
&  \includegraphics[scale=\graphletSZ]{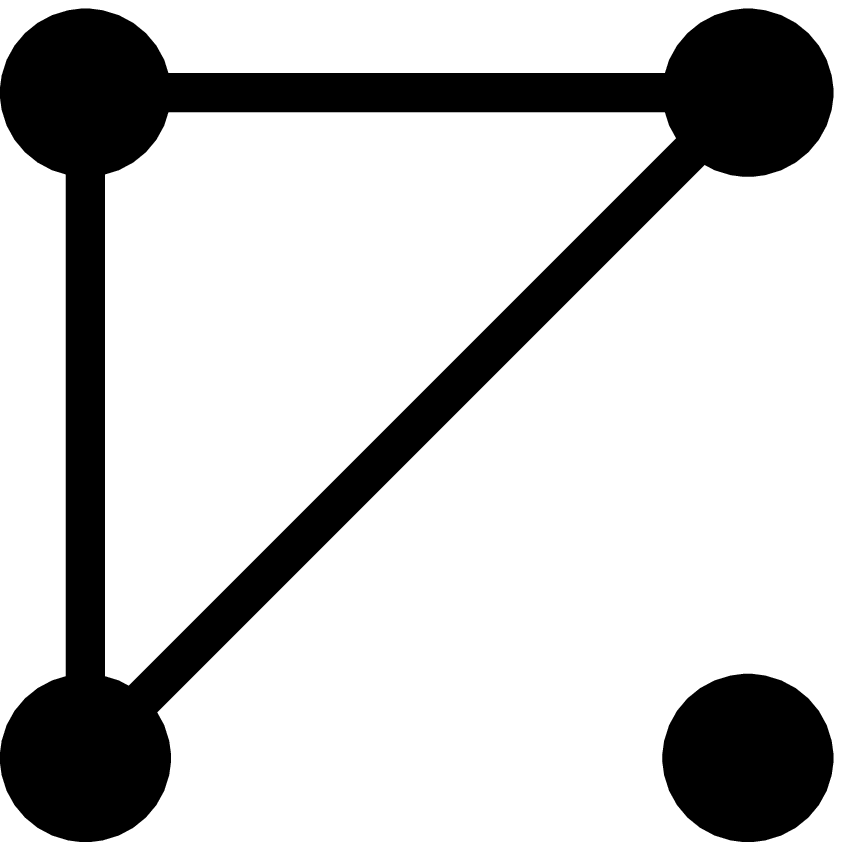} 
& $\g_{13}$  & 4-node-1-triangle 
\\

\TTT\BBB
& &
&   \includegraphics[scale=\graphletSZ]{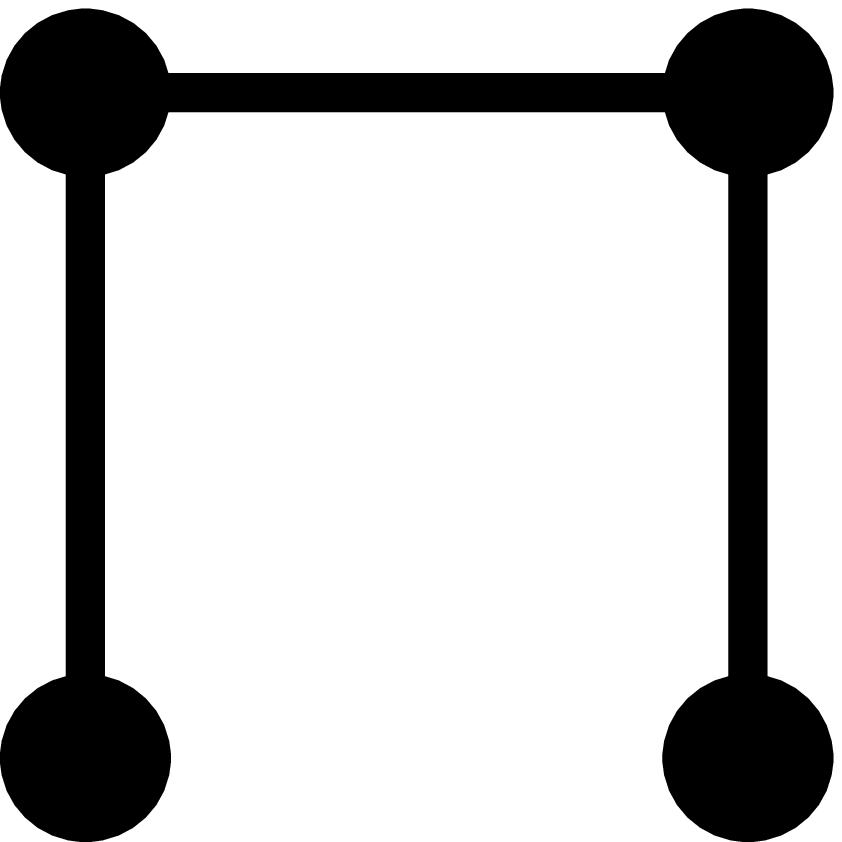} 
&  $\g_{12}$  & 4-path$^{{}_{\star}}$
\\
\bottomrule
\end{tabularx}
\end{table}
}

\section{Graphlet Decomposition} \label{sec:background}
\noindent
Graphlets are at the heart and foundation of many network analysis tasks (\eg, relational classification, network alignment)~\cite{prvzulj2004modeling,milenkoviae2008uncovering,hayes2013graphlet}.
Given the practical importance of graphlets, this paper proposes a hybrid CPU-GPU algorithm for computing the number of embeddings for both connected and disconnected k-vertex induced subgraphs (See Table~\ref{table:graphlet-notation}).

\subsection{Preliminaries}\label{sec:prelim}
\noindent
Let $G=(V,E)$ be an undirected graph where $V$ is the set of vertices and $E$ is its edge set.
The number of vertices is $n = |V|$ and number of edges is $m=|E|$.
We assume all vertex and edge sets are \textit{ordered}, i.e., $V = \{v_1,v_2,...,v_i,...,v_{\n}\}$ such that $v_{i-1}$ appears before $v_{i}$ and so forth.
Similarly, the ordered edges are denoted $E = \{e_1,e_2,...,e_i,...,e_{\m}\}$.
Given a vertex $v \in V$, let $\N(v) = \{w | (v,w) \in E\}$ be the set of vertices adjacent to $v$ in $G$. 
The degree $\d_{v}$ of $v \in V$ is the size of the neighborhood $|\N(v)|$ of $v$.
We also define $\dmax$ to be the largest degree in $G$.
\begin{mydef}[{\normalfont\bfseries Graphlet}\normalfont]\label{def:ind_graphlet} 
A graphlet $\g_i = (V_k,E_k)$ is a subgraph consisting of a subset $V_k \subset V$ of $k$ vertices from $G = (V,E)$ together with all edges whose endpoints are both in this subset $E_k = \{ \forall e \in E \,|\, e = (u,v) \wedge u,v \in V_k \}$.
\end{mydef}
\noindent
Let $\G^{(k)}$ denote the set of $k$-vertex induced subgraphs and $\G = \G^{(2)} \cup \cdots \cup \G^{(k)}$.
A $k$-graphlet is simply an \emph{induced subgraph} with k-vertices.

\subsection{Problem Formulation}
\noindent
It is important to distinguish between the two fundamental classes of graphlets, namely, \emph{connected} and \emph{disconnected} graphlets (see Table~\ref{table:graphlet-notation}). 
A graphlet is connected if there is a path from any node to any other node in the graphlet (see Definition~\ref{def:conn_graphlet}). 
Table~\ref{table:graphlet-notation} provides a summary of the connected and disconnected k-graphlets of size $k = \{2,3,4\}$.

\begin{mydef}[{\normalfont\bfseries Connected graphlet}\normalfont]\label{def:conn_graphlet}
A $k$-graphlet $\g_i=(V_k,E_k)$ is connected if there exists a path from any vertex to any other vertex in the graphlet $\g_i$, $ \forall u, v \in V_k,$\,$\exists P_{u-v}:u,\ldots,w,\ldots,v$, such that $d(u,v) \geq 0 \wedge d(u,v) \neq \infty$.
By definition, a connected graphlet $\g_i$ has only one connected component (\ie, $|C|=1$).
\end{mydef}

\begin{mydef}[{\normalfont\bfseries Disconnected graphlet}\normalfont]\label{def:disconn_graphlet}
A $k$-graphlet $\g_i=(V_k,E_k)$ is disconnected if there is not a path from any vertex $v \in \g_i$ to any other vertex $w \in \g_i$.
\end{mydef}

\noindent
Unlike most existing work that is only able to compute connected graphlets of a certain size (such as $k=4$), 
the goal of this work is to compute the frequency of both connected and disconnected graphlets of size $k \in \{2,3,4\}$.
More formally, 

\begin{myproblem}[{\normalfont\bfseries Global graphlet counting}\normalfont]\label{prob:macro}
Given the graph $G$, find the number of embeddings (appearances) of each graphlet $\g_i \in \G$ in the input graph $G$.
We refer to this problem as the \emph{global graphlet counting problem}. 
A graphlet $\g_i \in \G$ is embedded in $G$, iff there is an injective mapping $\sigma: V_{\g_i} \rightarrow V$, with $e=(u,v) \in E_{\g_i}$ if and only if $e'=(\sigma(u),\sigma(v)) \in E$. 
\end{myproblem}

\section{Hybrid CPU-GPU Framework} \label{sec:hybrid-framework}
\noindent
This work proposes parallel graphlet decomposition methods that are designed to leverage 
(i) parallelism (multiple cores on a CPU or GPU) as well as 
(ii) heterogeneity that leverages simultaneous use of a CPU and GPU (as well as multiple CPUs and GPUs).
To the best of our knowledge, this is the first work to use multiple GPUs (and of course multiple GPUs and CPUs) for computing induced subgraph statistics.
\smallskip
\begin{compactitem}
\item Single GPU Methods (using multiple cores)

\item Multi-GPU Methods. 

\item Hybrid Multi-core CPU-GPU Methods.
\end{compactitem}
\medskip

\noindent
Methods from all three classes are shown to be effective on a large collection of graphs from a variety of domains (e.g., biological, social, and information networks~\cite{nr-aaai15}).
In particular, methods from these classes have three important benefits.
First, the performance is orders of magnitude faster than the state-of-the-art.
Second, the GPU methods are cost effective enjoying superior performance per capita.
Third, the performance per watt is significantly better than existing traditional CPU methods.

\begin{figure}[t!]
\centering
\includegraphics[width=0.9\linewidth]{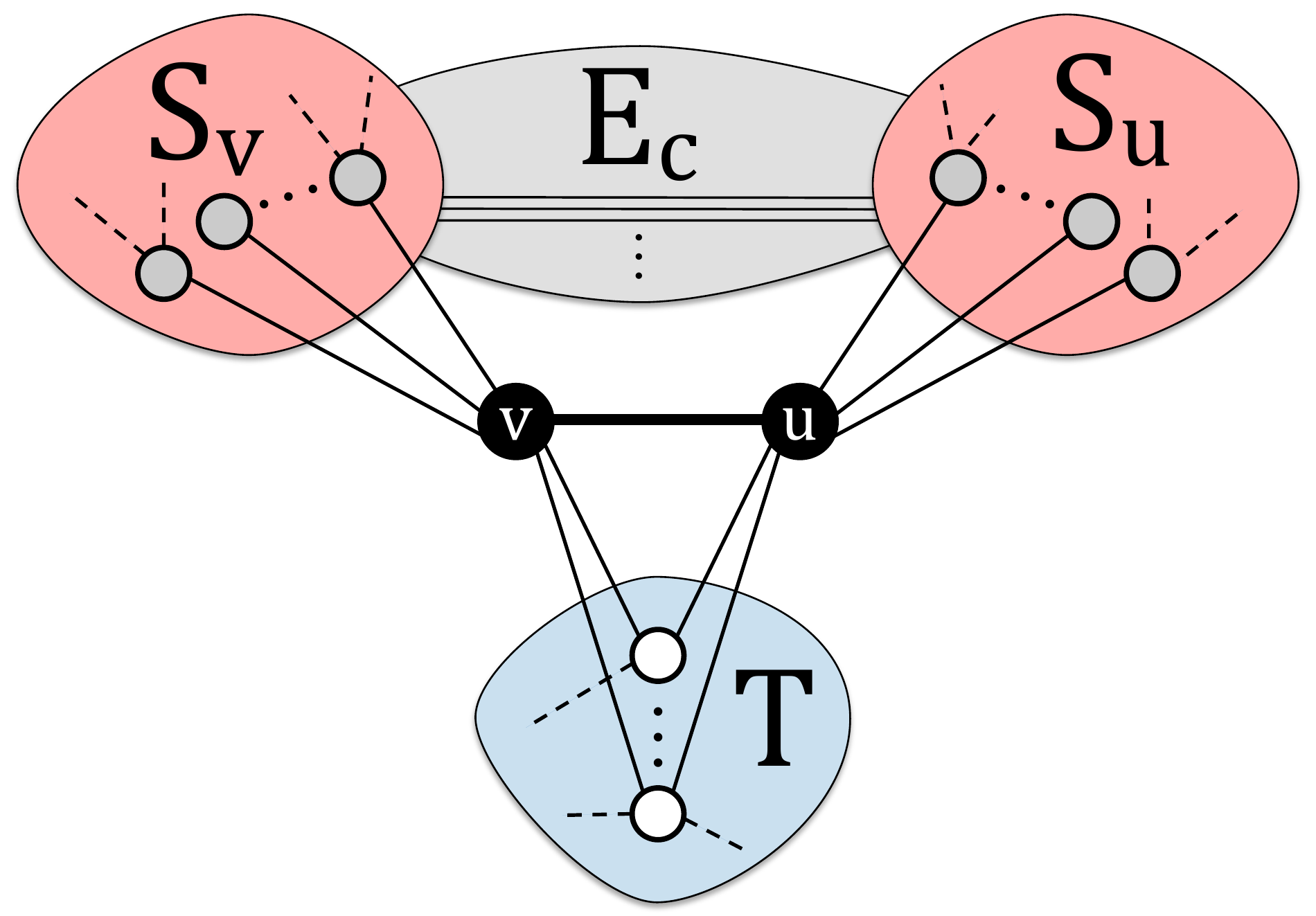}
\vspace{-2mm}
\caption{Let $\tri$ be the set of nodes completing a triangle with the edge $(v,u) \in \E$, and let $\S_v$ and $\S_u$ be the set of nodes that form a 2-star with $v$ and $u$, respectively.
Note that $\S_u \cap \S_v = \emptyset$ by construction and $|\S_u \cup \S_v| = |\S_u|+|\S_v|$.
Further, let $\E_c$ be the set of edges that complete a cycle (of size 4) $\wrt$ the edge $(v,u)$ where for each edge $(p,q) \in \E_c$ such that $p \in \S_v$ and $q \in \S_u$ and both $(p \cap \S_u) \cup (q \cap \S_v) = \emptyset$, that is, $p$ is not adjacent to $u$ ($p \not\in \N(u)$) and $q$ is not adjacent to $v$ ($q \not\in \N(v)$).
}
\label{fig:graphlets-intuition}
\end{figure}

\subsection{Parallel Graphlet Computations}  \label{sec:parallel-overview}
\noindent
Our algorithm searches over the set of edges $E$ of the input graph $G=(V,E)$.
Given an edge $\e=(u,v) \in E$, let $\N(\e)$ denote the edge neighborhood of $\e$ defined as:
\begin{equation}\label{eq:edge-neighborhood}
\N(e) = \N(u,v) = \{\N(u) \cup \N(v) \setminus \{u,v\}\},
\end{equation} 
\noindent
where $\N(u) \textrm{ and } \N(v)$ are the neighbors of $u \textrm{ and } v$, respectively.
For convenience, let $\Ng_{\e} = G(\{\N(v) - u\} \cup \{\N(u) - v\})$ be the (explicit) edge-induced neighborhood subgraph.
Given an edge $e=(u,v) \in E$, we explore the subgraph surrounding $\e$ (called the egonet of $e$), \ie, the subgraph induced by both its endpoints and the vertices in its neighborhood. 
Our approach uses a specialized graph encoding based on the edge-CSC representation~\cite{rossi2014tcore}.

The parallel scheme leverages the fact that the induced subgraph (graphlet) problem can be solved independently for each edge-centric neighborhood $\N(\e_i) \in \{\N(e_1), \ldots, \N(e_{\m})\}$ in $G$, and therefore may be computed simultaneously in parallel.
A processing unit denoted by $\omega$ refers to a single CPU/GPU worker (core).
In the context of message-passing and distributed memory parallel computing, a node refers to another machine on the network with its own set of multi-core CPUs, GPUs, and memory.
Other important properties include 
the search order $\Pi$ in which edges are solved in parallel, 
the batch size $\b$ (number of jobs/tasks assigned to a worker by a dynamic scheduling routine), 
and the dynamic assignment of jobs (for load balancing).

While there are only a few such parallel graphlet algorithms, with the exception of $\pgd$ all of these methods are based on searching over the vertices (as opposed to the edges).
However, as we shall see, the parallel performance of these approaches are guaranteed to suffer more from load balancing issues, communication costs, and other issues such as curse of the last reducer, etc.

\smallskip
\noindent
\textbf{Improved Load Balancing}:
Let $\vz_j \in \RR^{\n}$ and $\vx_j \in \RR^{\m}$ be counts of an arbitrary graphlet $\g_j \in \G$ for vertices $\n=|V|$ and edges $\m=|E|$, respectively.
Given a vertex $v_i \in V$ and an edge $\e_k \in E$, 
let $z^{}_{ij}$ and $x^{}_{kj}$ denote the number of vertex and edge incident counts of a graphlet $\g_j$ for vertex $v_i$ and edge $\e_k$, respectively.
Furthermore, let 
\begin{equation} \label{eq:graphlet-count-equal}
Z_{j} = \sum_{v_i \in V} z_{ij} \;\text{   and   }\; X_{j} = \sum_{e_k \in \E} x_{kj}
\end{equation}
\noindent 
where $Z_j$ and $X_j$ are the global frequency of graphlet $\g_j$ in the graph $G$.
Thus, it is straightforward to verify that $Z_j = X_j$.
Further, let $\bar{Z}_{j} = Z_{j} / \n$ and $\bar{X}_{j} = X_j/\m$ 
be the mean vertex and edge count for graphlet $\g_j \in \G$, respectively.
Now, assuming $\n \ll \m$,\footnote{This holds in practice for nearly all real-world graphs} then $\bar{X}_{j} < \bar{Z}_{j}$.
Clearly, more work is required to compute graphlets for each vertex $v_i$ on average (compared to the number of graphlets counted per edge).
This implies that edge-centric parallel algorithms are guaranteed to have better load balancing (among other important advantages) than existing vertex-centric algorithms.

\subsection{Preprocessing Steps} \label{sec:preprocessing}
\noindent
Our approach benefits from the preprocessing steps below and the useful computational properties that arise.

\begin{myitemize}
\smallskip
\item[$\mathbf{P1}$] The vertices $V=\{v_1,\ldots,v_{\n}\}$ are sorted from smallest to largest degree and relabeled such that $\d(v_1) \leq \d(v_2) \leq \d(v_i) \leq \d(v_{\n})$.

\item[$\mathbf{P2}$] 
For each $\N(v_i) \in \{ \N(v_1), \ldots, \N(v_{\n}) \}$, order the neighbors $\N(v_i) = \{\ldots,w_j,\ldots,w_k,\ldots\}$ s.t. $j<k$ if $f(w_j) \geq f(w_k)$.
Thus, the set of neighbors $\N(v_i)$ are ordered from largest to smallest degree.

\item[$\mathbf{P3}$]
Given an edge $(v,u) \in \E$, we ensure that $v$ is always the vertex with largest degree $\d_v$, that is, $\d_v \geq \d_u$.
This gives rise to many useful properties and as we shall see can lead to a significant reduction in runtime.
For instance, our approach avoids searching both $\S_v$ and $\S_u$ for computing 4-cycles, and instead, allows us to compute 4-cycles by simply searching one of those sets.
Thus, our approach always computes 4-cycles using $\S_u$ since (by the property above) is guaranteed to be less work/faster than if $\S_v$ is used (and the runtime difference can be quite significant).

\end{myitemize}
\noindent
Each step above is computed in $\O(\n)$ or $\O(\m)$ time and is easily parallelized.

\subsection{Hybrid CPU-GPU}
\label{sec:hybrid-algorithm}
\noindent
The algorithm begins by computing an edge ordering $\Pi=\{ \e_1, \ldots, \e_{\m} \}$ where the edges that are most difficult (with highly skewed, irregular, unbalanced degrees) are placed upfront, followed by edges that are more evenly balanced.
In other words, $\Pi$ is a permutation of the edges by some function or graph property $f(\cdot)$ such that $k<j$ for $e_k$ and $e_j$ if $f(e_k) > f(e_j)$, and ties are broken arbitrarily (\eg, using ids).
For instance, edges can be ordered from largest to smallest degree (a proxy for the difficulty and unbalanced nature of the edge).
For implementation purposes, $\Pi$ is essentially a double-ended queue (dequeue) where elements can be added or removed from either end (\ie, push and pop operations at either end).
Afterwards, the previous edge ordering $\Pi$ is split into three \emph{initial} sets:
\begin{equation}\label{eq:double-ended-queue}
\Pi = \big\{
\underbrace{\e_1, \ldots, \e_{k}}_{\Pi_{\rm cpu}}, 
\overbracket[0.8pt][8pt]
{\underbrace{\e_{k+1},\ldots,\e_{j-1}}_{\Pi_{\rm unproc}}}^{{\rm unprocessed}\, (j-k)-1},
\underbrace{\e_{j}, \e_{j+1}, \ldots, \e_{\m}}_{\Pi_{\rm gpu}}
\big\}
\end{equation}

\noindent
Now, the edges $\Pi_{\rm gpu} = \{\e_{j}, \e_{j+1}, \e_{j+2}, \ldots, \e_{\m} \}$\footnote{The edges can be thought of as edge neighborhoods, since the induced subgraphs are counted over each edge neighborhood.} are split into $\ngpu$ disjoint sets $\{ \I_1, \I_2, \ldots , \I_{\ngpu} \}$ with approximately equal work among each GPU device. This is accomplished by partitioning the edges in a round-robin fashion.
Each GPU computes the induced subgraphs centered at each of the edges in $\I_{i}$, which can be thought of as a local job queue for a particular GPU. 
Similarly, the CPU workers compute the induced subgraphs centered at each edge in $\Pi_{\rm cpu}$. 
Once an edge is assigned, it is removed from the corresponding local queue (for either CPUs or GPUs).

Once a CPU worker finishes all edges (or more generally tasks) in its local queue, 
it takes (dequeues) the next $b$ unprocessed edges from the \emph{front} of $\Pi_{\rm unproc}$ (the global queue of remaining/unassigned work) and pushes them to its local queue.
On the other hand, once a GPU's local queue becomes empty (and thus the GPU becomes idle), it is assigned the next chunk of unprocessed edges from the \emph{back} of the queue (\ie, dequeued and pushed onto that GPU's local queue).
Unlike the CPU, we must transfer the assigned edges to the corresponding GPU.

Finally, the graphlet counts from the multiple CPUs and GPU devices are combined to arrive at the final global and local graphlet counts.
Similarly, if the local graphlet counts for each edge are warranted (also known as micro graphlet counts), then one would simply combine the per edge results to arrive at the final counts for each edge (and each graphlet).
Recall that to compute all $k\in\{2,3,4\}$-vertex graphlets, our algorithm only requires us to store counts for triangles, cliques, and cycles, and from these, we can easily derive the other counts for both connected and disconnected graphlets in constant time.
This not only avoids communication costs (and thus significantly reduces the amount of communications that would otherwise be needed by other algorithms), but also reduces space and time.

{
\smallskip\noindent
Other important aspects include:
\begin{itemize}
\item 
Dynamic load balancing is performed for both CPUs and GPUs. 
For the CPUs, once a worker completes the tasks in its local queue, it immediately takes the next $b$ tasks from the front of $\Pi$. 
We typically use a very small chunk size $b$ for the CPU.
The intuition is that these tasks are likely skewed and thus may take a significant amount of time to complete.
Moreover, the overhead associated with this dynamic load balancing on the CPU is quite small (relative to the GPU of course, where the communication costs are significantly larger).

\item To avoid communication costs and other performance degrading behavior, it is important that the GPUs are initially assigned a large fraction of the edges to process (which are then split among the GPUs).
For majority of large real-world networks (with power-law), GPUs are initially assigned about $80\%$ of the edges, and this seemed to work well, as it avoids both extremes (that is, significantly under- or oversubscribing the GPUs).
In particular, significantly oversubscribing the GPUs causes a lot of work to be stolen by the GPUs (and the overhead associated with it, such as additional communication costs for the edges that will be stolen by the CPU (or even another GPU), whereas significantly undersubscribing the GPUs increase the load balancing overhead (causing additional communication costs, etc.).
Ideally, we would want to assign the largest fraction of edges to the GPUs such that both the GPUs and CPUs finish at exactly the same time, and thus, avoid any communication or other costs associated with load balancing, etc.

\item 
Note $b$ is a chunk size, and is different for CPU and GPU devices. 
In particular, for CPU, we typically set $b=1$, since these tasks are the most difficult to compute, and the runtime for each is likely to be skewed and irregular. This also helps avoid costs associated with work-stealing, which occurs when all such edge-centric tasks have been assigned, but not yet finished.
Hence, to avoid the case where all but a single CPU workers have finished, and the remaining CPU worker has many edge-centric tasks in its local queue.
In this case, of course, the tasks remaining in that CPU workers local queue would be stolen and distributed among the idle CPU workers.
Note that as discussed later, one can also divide tasks at a much finer level of granularity, as many of the core computations carried out for a single edge-centric task can be computed independently.
For instance, cliques and cycles are completely independent once the sets $\tri$ and $\S_u$ are computed.

\item Shared job queue with \emph{work stealing} so that every multi-core CPU and GPU remains fully utilized.

\item Data is never moved once it is partitioned and distributed to the workers.

\item It is straightforward to see that $|T_u| \leq |\N(u)|$, that is, the number of triangles centered at u is bounded above by the number of neighbors (degree of u).
Similarly, $|S_u| \leq |N(u)|$, that is, the number of $2$-stars centered at $u$ is bounded above by $|\N(u)|$.
Hence, $|S_u| + |T_u| = |\N(u)|$.
See Figure~\ref{fig:graphlets-intuition} for further intuition.
We use the above fact to reduce the space requirements as well as improve locality.

\item 
Given two vertices $w_i, w_j \in \tri$ that form a triangle with $\e_k = (v,u)$\footnote{Thus, there are two triangles centered at $\e_k=(v,u)$, namely, $\{v,u,w_i\}$ and $\{v,u,w_j\}$.}, what is the most efficient search strategy (Alg.~\ref{alg:cliques-res})? 
As a result of the past ordering and relabeling, searching the neighbors of $w_j$ for $w_i$ always results in less work.
Observe that since the vertices in the set $\tri$ are ordered from largest to smallest degree, then $|\N(w_j)| < |\N(w_i)|$, and thus, we search $|\N(w_j)|$ for vertex $w_i$.
Hence, if $(w_i,w_j) \in E$, then the subgraph induced by $\{v,u,w_i,w_j\}$ denoted $G(\{v,u,w_i,w_j\})$ is a 4-clique $\g_7$ (See Table~\ref{table:graphlet-notation})
\end{itemize}

{
\algblockdefx[parallel]{ParFor}{EndPar}[1][]{$\textbf{parallel for}$ #1 $\textbf{do}$}{$\textbf{end parallel}$}
\algrenewcommand{\alglinenumber}[1]{\fontsize{7}{8}\selectfont#1}
\algtext*{EndPar}
\begin{figure}[t!]
\begin{center}
\begin{minipage}{1.0\linewidth}
\begin{algorithm}[H]
\caption{\,\small {This algorithm computes all 3-vertex graphlets via neighbor iteration and also leverages a hash table for fast lookups. 
Used mainly for CPU-based enumeration of 3-vertex induced subgraphs.
}}
\label{alg:3-graphlets-hash-neigh-iter}
\begin{spacing}{1.32}
\fontsize{8}{9}\selectfont
\begin{algorithmic}[1]
\Ensure~\newline
\quad\quad A set $\tri$ of vertices that complete triangles with $\e_k$
\newline
\quad\quad A set $\S_u$ of vertices that form 2-stars centered at $u$ with $\e_k$
\smallskip
\Procedure {3-Graphlets}{$G, \e_k, \hash$}

\State Set $\tri \leftarrow \emptyset$ and $\S_u \leftarrow \emptyset$ \label{algline:3-graphlet-init}
\ParFor [$w \in \N(v)$ {\bf where} $w \not= u$] \label{algline:graphlet-create-hash}
	\State Set $\hash(w) = \lambda_1$ \label{algline:graphlet-mark-neigh-of-v}
\EndPar

\ParFor [$w \in \N(u)$ {\bf where} $w \not= v$] \label{algline:triangles-and-wedges}
	\If{$\hash(w) = \lambda_1$} \label{algline:triangle} 
		\State $\tri \leftarrow \tri \cup \{w\}$ and set $\hash(w)=\lambda_3$ \label{algline:triangle-found} \Comment{triangle}
	\Else \, $\S_u \leftarrow \S_u \cup \{w\}$ and set $\hash(w)=\lambda_2$ \label{algline:2star-found} \Comment{2-star} 
	\EndIf
\EndPar
\State Set $\X_{k,3} \leftarrow |\tri|$
\EndProcedure
\smallskip
\end{algorithmic}
\end{spacing}
\end{algorithm}
\end{minipage}
\end{center}
\vspace{-4.mm}
\end{figure}
}

{
\algblockdefx[parallel]{ParallelFor}{EndParallel}[1][]{$\textbf{parallel for}$ #1 $\textbf{do}$}{$\textbf{end parallel}$}
\algrenewcommand{\alglinenumber}[1]{\fontsize{7}{8}\selectfont#1}
\begin{figure}[t!]
\vspace{-5.mm}
\begin{center}
\begin{minipage}{1.0\linewidth}
\begin{algorithm}[H]
\caption{\,\small
{This algorithm computes all \text{3-vertex graphlets} using \emph{binary search}.
For multi-core architectures with limited memory (such as GPUs).
}}
\label{alg:3-graphlets-bsearch}
\begin{spacing}{1.32}
\fontsize{8}{9}\selectfont
\begin{algorithmic}[1]
\Procedure {3-Graphlets-BinSearch}
{$G$, $\e_k$}

\State Let $\e_k = (v,u) \in \E$ be an edge in $G$ \emph{s.t.} $\d_v \geq \d_u$
\State Set $\tri \leftarrow \emptyset$ and $\S_u \leftarrow \emptyset$ \label{algline:3-graphlet-init}

\ParallelFor[{\bf each} $w \in \N(u)$ {\bf where} $w \not= v$]		\label{algline:T-and-Su} 
		\If{$v \in \N(w)$ via $\mathsf{binary}$ $\mathsf{search}$} \label{algline:found-tri-bsearch} \Comment{$\O(\log \d_w)$}
			\State \,\, $\tri \leftarrow \tri \cup \{w\}$ \label{algline:add-vertex-to-T} \Comment{triangle} 
		\Else
			\, $\S_u \leftarrow \S_u \cup \{w\}$ \label{algline:add-vertex-to-Su} \Comment{$2$-star} 
		\EndIf
\EndParallel
\State Set $\X_{k,3} \leftarrow |\tri|$
\EndProcedure
\smallskip
\end{algorithmic}
\end{spacing}
\end{algorithm}
\end{minipage}
\end{center}
\vspace{-6.mm}
\end{figure}
}

\noindent
More importantly, our hybrid multi-core algorithm ensures all CPUs and GPU devices are always working simultaneously.
The only exception arises when all but a single edge remains (is currently being processed by one or more other workers).
In this case, there is no further work remaining (as the edge task has already been split and assigned to other workers).
Nevertheless, we stress that this case is unlikely due to the fact that edges with large degrees are always processed first by both GPU and CPU workers.
Thus, the remaining edge is likely to be easily computed.
As mentioned, even a single edge neighborhood has many independent components that can be computed in parallel by other workers.
Work-stealing is used to ensure that all processing units are fully utilized.

\subsection{Finer-Granularity and Work Stealing} \label{sec:finer-granularity-and-work-stealing}
\noindent
Each job in $\Pi$ may represent the computations required for a single edge, or they may represent an even smaller unit of work.
For a single edge $\e_k \in \E$, we compute
(i) the sets $\tri$ and $S_u$, then 
(ii) we find the total k-cliques for a given edge using $\tri$, and 
(iii) the total cycles of size $k$ using $S_u$. 
Note that (ii) and (iii) are independent and thus can be computed in parallel.
A job may also represent these smaller units of work.
For instance, if $T$ (or $S_u$) are large and computationally expensive to compute, then the worker in the simplest case can push $T$ (or $S_u$) on the job queue for others to work on.
Further, one can split $T$ (or $\S_u$) into even more finer grained independent jobs such that:
\begin{align}
T = \{
\underbrace{w_1, \cdots, w_{i}}_{T_{1:i}}, 
\underbrace{w_{i+1}, \cdots, w_t}_{T_{i+1:t}}
\}
\end{align}
\noindent
Whenever possible work is stolen locally to minimize communication. 
Notice that if work is stolen locally from a GPU, \ie, a GPU worker computes $4$-cliques for a given edge $\e_k$ using the already computed set $\tri$, then expensive communication is avoided due to the set $\tri$ being stored in global memory.
In fact, one could store $\tri$ and $\S_u$ contiguously in a single global array, and since $|\tri|+|\S_u|=\d_u$, then we can allocate contiguous memory for storing such sets for each edge.
Alternatively, one could simply allocate an array of size $\dmax \cdot P_i$ where $P_i$ is the number of cores (unique workers) for the $i^{\th}$ GPU (or CPU).
Hence, $\dmax$ serves as an upper bound and in most cases requires significantly less space than the previous approach, since each GPU worker simply indexes into their own subarray which stores $\tri$ and $\S_u$.
However, one may need to communicate $\tri$ so that the GPU worker to compute $4$-cliques from $\tri$ has a copy (stored in their subarray).
The only exception is if that is the last edge to be assigned to that GPU, and thus, the state of the subarray can persist until termination.
However, suppose work is stolen from another GPU (not locally), then we would need to communicate not only the edge id, but the portion of $\tri$ (or $\S_u$) assigned to the GPU worker would also be needed.

To avoid bulk synchronization, workers store and aggregate statistics locally, and thus avoid unnecessary communications.
In the case of global graphlet statistics computed for $G$, each worker maintains local aggregates and communicates them only once upon completion and thus has a cost of $O(\kappa)$ where $\kappa$ is the number of induced subgraph statistics.

{
\algrenewcommand{\alglinenumber}[1]{\fontsize{7}{8}\selectfont#1}
\begin{figure}[t!]
\begin{center}
\begin{minipage}{1.0\linewidth}
\begin{algorithm}[H]
\caption{\,{
Clique counts restricted to searching $\tri$
}}
\label{alg:cliques-res}
\begin{spacing}{1.24}
\fontsize{8}{9}\selectfont
\begin{algorithmic}[1]
\Procedure {CliqueRes}{
$\tri$, $\e_k$
}
\State Set $\X_{k,7} \leftarrow 0$
\parfor[{\bf each} $w_i \in \tri$ in {\bf order} $w_1, w_2, \cdots$ of $\tri$]
\label{algline:clique-res-wi}
	\ForAll{{\bf each} $w_j \in \{w_{i+1}, ..., w_{|\tri|} \}$ \textsf{in order}} \label{algline:clique-res-wj}
		\If{$w_i \in \N(w_j)$ via $\mathsf{binary}$ $\mathsf{search}$} \label{algline:clique-res-bsearch}
			\State $\X_{k,7} \leftarrow \X_{k,7} + 1$ 
		\EndIf
	\EndFor
\endpar
\State \textbf{return} $\X_{k,7}$
\EndProcedure
\smallskip
\end{algorithmic}
\end{spacing}
\end{algorithm}
\vspace{-8.8mm}
\begin{algorithm}[H]
\caption{\,{%
Cycle counts restricted to $\S_u$ and $\S_v$
}}
\label{alg:cycles-res}
\begin{spacing}{1.24}
\fontsize{8}{9}\selectfont
\begin{algorithmic}[1]
\Procedure {CycleRes}{
$\S_u$, $\S_v$, $\e_k$
}
\State Set $\X_{k,10} \leftarrow 0$
\parfor[{\bf each} $w \in \S_u$]		\label{algline:cycle-res-Su}
	\ForAll {$r \in \S_v$} 	\label{algline:cycle-res-Sv}
	
		\If{$r \in \N(w)$ via $\mathsf{binary}$ $\mathsf{search}$} \label{algline:cycle-res-bsearch}
			\State 
			$\X_{k,10}\leftarrow \X_{k,10} + 1$ 
		\EndIf
	\EndFor
\endpar
\State \textbf{return} $\X_{k,10}$ 
\EndProcedure
\smallskip
\end{algorithmic}
\end{spacing}
\end{algorithm}
\end{minipage}
\end{center}
\end{figure}
}

{
\algblockdefx[parallel]{ParFor}{EndPar}[1][]{$\textbf{parallel for}$ #1 $\textbf{do}$}{$\textbf{end parallel}$}
\algrenewcommand{\alglinenumber}[1]{\fontsize{7}{8}\selectfont#1}
\begin{figure}[b!]
\begin{center}
\begin{minipage}{1.0\linewidth}
\begin{algorithm}[H]
\caption{\,{Clique counts via neighbor iteration
}}
\label{alg:cliques}
\begin{spacing}{1.22}
\fontsize{8}{9}\selectfont
\begin{algorithmic}[1]
\Procedure {Clique}{
$\hash$, 
$\tri$, 
$\e_k$}
\State Set $\X_{k,7} \leftarrow 0$ \label{algline:clique-init}
\ParFor[{\bf each} $w \in \tri$] \label{algline:clique-outer-parfor-T}
	\For{{\bf each} $r \in \N(w)$} \label{algline:clique-inner-for-neigh}
		\If{$\hash(r) = \lambda_3$} Set $\X_{k,7} \leftarrow \X_{k,7} + 1$ \eat{\Comment{found $4$-Clique}} \label{algline:clique-detected}		
		\EndIf
	\EndFor
	\State Reset $\hash(w)$ to $0$  \label{algline:clique-reset-hash}	
\EndPar
\State \textbf{return} $\X_{k,7}$ \label{algline:clique-return}
\EndProcedure
\smallskip
\end{algorithmic}
\end{spacing}
\end{algorithm}
\vspace{-8.8mm}
\begin{algorithm}[H]
\caption{\,{Cycle counts via neighbor iteration
}}
\label{alg:cycles}
\begin{spacing}{1.22}
\fontsize{8}{9}\selectfont
\begin{algorithmic}[1]
\Procedure {Cycle}{
$\hash$, 
$\S_u$,
$\e_k$} 
\State Set $\X_{k,10} \leftarrow 0$ \label{algline:cycle-init}
\ParFor[{\bf each} $w \in \S_u$] \label{algline:cycle-outer-parfor-Su}
	\For{{\bf each} $r \in \N(w)$} \If{$\hash(r) = \lambda_2$} set $\X_{k,10} \leftarrow \X_{k,10} + 1$ \eat{\Comment{found $4$-Cycle}} \label{algline:cycle-detected}
		\EndIf
	\EndFor
	\State Reset $\hash(w)$ to $0$ \label{algline:cycle-reset-hash}
\EndPar
\State \textbf{return} $\X_{k,10}$  \label{algline:cycle-return}
\EndProcedure
\smallskip
\end{algorithmic}
\end{spacing}
\end{algorithm}
\end{minipage}
\end{center}
\vspace{-4.mm}
\end{figure}
}

\subsection{Unrestricted counts} \label{sec:unrestricted-counts}\noindent
\noindent
The GPU workers use binary search to derive the sets $\tri$ and $\S_u$ from Alg.~\ref{alg:3-graphlets-bsearch}, whereas 
the CPU workers compute $\tri$ and $\S_u$ from Alg.~\ref{alg:3-graphlets-hash-neigh-iter} (or Alg.~\ref{alg:3-graphlets-bsearch} if memory is limited and/or dynamically selected).
The key difference is that the CPU workers create a fast hash table on the neighbors of $v$ for $\e_k=(v,u)$ allowing for $o(1)$ time checks.
Moreover, the hash table is also exploited for encoding nodes with particular types and enabling us to check an arbitrary type of vertex in $o(1)$ time.
Note that the $\lambda_i$'s in Alg.~\ref{alg:3-graphlets-hash-neigh-iter} represent distinct vertex types, and are used later for finding cliques (Alg.~\ref{alg:cliques}) and cycles (Alg.~\ref{alg:cycles}) extremely fast.
We also avoid the $O(|\N(v)|)$ time it costs to reset $\hash$ for each edge by defining $\lambda_i$ to be unique for each edge.
Observe that each CPU worker in Alg.~\ref{alg:3-graphlets-hash-neigh-iter} maintains $\hash$ taking $\O(\n)$ space, whereas Alg.~\ref{alg:3-graphlets-bsearch} only requires $\O(|\tri| + |\S_u|) = \O(\d_u)$ space to store $\tri$ and $\S_u$ (which Alg.~\ref{alg:3-graphlets-hash-neigh-iter} also uses).
Note that $\S_v$ is easily computed (if needed) from Alg.~\ref{alg:3-graphlets-hash-neigh-iter} and Alg.~\ref{alg:3-graphlets-bsearch} by simply setting $\S_v \leftarrow \N(v)$, and then removing each vertex $w \in \tri_{e}$ (on-the-fly), that is, $\S_v \leftarrow \S_v \setminus \{w\}$.

The proposed approach derives all $k$-graphlets for $k\in\{2,3,4\}$ using only the local edge-based counts of triangles, cliques, and cycles, along with a few other constant time graph and vertex parameters such as number of vertices $\n=|V|$, edges $\m=|E|$, as well as vertex degree $\d_{v}=|\N(v)|$.
Given an edge $\e_k \in E$ from $G$,
let $\X_{k,3}$, $\X_{k,7}$, and $\X_{k,10}$ be the frequency of triangles, cliques, and cycles centered at the edge $e_k \in E$ in the graph $G$, respectively.
Observe that $\X_{k,i}$ (or simply $x_i$) is the count of the induced subgraph $\g_i$ for an arbitrary edge $\e_k$ (See Table~\ref{table:graphlet-notation}).
The local (micro-level) $3$-graphlets for edge $\e_k$ are as follows:
\begin{align}
&x_{3} = |\tri| \\
&x_{4} = (\d_u + \d_v - 2) - 2|\tri| \\
&x_{5} = \n - x_{4} + |\tri| - 2 \\
&x_{6} = \mychoose{\n}{3} - (x_{3} + x_{4} + x_{5})
\end{align}
\noindent
Further, notice that given $x_3 = |\tri|$ for $\e_{k}=(v,u) \in \E$, 
we can derive $|\S_u|$ and $|\S_v|$ (that is, the number of 2-star patterns centered at $u$ and $v$ of $\e_k$, respectively) as:
\begin{align}
& |\S_u| = \d_u - |\tri| - 1 \\
& |\S_v| = \d_v - |\tri| - 1
\end{align}
\noindent
Therefore, the number of two-stars centered at $\e_k$ denoted $x_4$ can be rewritten simply as $x_4 = |\S_u| + |\S_v|$.
These $3$-vertex induced subgraph statistics are then used as a basis to derive the induced subgraphs of size $k+1$.

Notice that GPU workers compute $\X_{k,3}$ (as well as $\tri$ and $\S_u$) for edge $\e_k$ using Alg.~\ref{alg:3-graphlets-bsearch}, whereas CPU workers compute $\X_{k,3}$ from  Alg.~\ref{alg:3-graphlets-hash-neigh-iter}.
Afterwards, $\X_{k,7}$ and $\X_{k,10}$ can be computed in any order as they are completely independent, and can even be stolen by another parallel worker (CPU and/or GPU worker that requires more work).
A GPU worker computes the number of 4-cliques $\X_{k,7}$ centered at edge $\e_k$ via Alg.~\ref{alg:cliques-res}, whereas a CPU worker mainly leverages Alg.~\ref{alg:cliques} for computing $\X_{k,7}$, but may also exploit Alg.~\ref{alg:cliques-res} if determined (dynamically) that it requires less work than the other approach.
Similarly, a GPU worker computes the number of 4-cycles $\X_{k,10}$ centered at $\e_k$ via Alg.~\ref{alg:cycles-res}, whereas a CPU worker computes $\X_{k,10}$ via Alg.~\ref{alg:cycles}.

Given only the triangles $\X_{k,3}$, cliques $\X_{k,7}$, and cycles $\X_{k,10}$ for each edge $\e_{k}=(v,u) \in \E$, 
we derive the \emph{unrestricted counts} for \emph{connected} and \emph{disconnected graphlets} of size $k \in \{3,4\}$.
We first derive the unrestricted counts for \emph{connected and disconnected 3-graphlets} as follows: 
\begin{align}
& \C_{3}    = \sum_{\e_{k}=(v,u) \in \E} \X_{k,3} \quad\, = \sum_{\e_{k}=(v,u) \in \E} |\tri| \\
& \C_{4}		= \sum_{\e_{k}=(v,u) \in \E} |\S_v| + |\S_u| \\
& \C_{5} 	= \sum_{\e_{k}=(v,u) \in \E} \n - (|\S_v|+|\S_u| + |\tri|) - 2  
\end{align}
\noindent
Note that $\C_{6}$ is not needed since $X_6$ can be computed directly from $X_3$, $X_{4}$, and $X_{5}$.
The \emph{unrestricted counts} for the \emph{connected 4-graphlets} (4-vertex connected induced subgraphs) are derived as follows:
\begin{align}
& \C_{7} = \sum_{\e_{k}=(v,u) \in \E} \X_{k,7} \\ 
& \C_{8} = \sum_{\e_{k}=(v,u) \in \E} \mychoose{\tri}{2} \\ 
& \C_{9} 	= \sum_{\e_{k}=(v,u) \in \E} |\tri| \cdot |\S_v| \cdot |\S_u| \\  
& \C_{10} 	= \sum_{\e_{k}=(v,u) \in \E} \X_{k,10} \\ 
& \C_{11} 	= \sum_{\e_{k}=(v,u) \in \E}  \mychoose{|\S_v|}{2} + \mychoose{|\S_u|}{2} \\
& \C_{12} 	= \sum_{\e_{k}=(v,u) \in \E} |S_v|\cdot|S_u|  
\end{align}
\noindent
Given an arbitrary edge $\e_k \in \E$, we define $D_{\e} = \n - (|\S_v| + |\S_u| + |\tri|) - 2$ for convenience.
The \emph{unrestricted counts} for the \emph{disconnected 4-graphlets} (4-vertex disconnected induced subgraphs) are derived as follows:
\begin{align}
& \C_{13} 	= \sum_{\e_{k}=(v,u) \in \E} |\tri| \cdot D_{\e} \\ 
& \C_{14} 	= \sum_{\e_{k}=(v,u) \in \E} \m - \d_v - \d_u + 1 \\
& \C_{15} 	= \sum_{\e_{k}=(v,u) \in \E}  (|\S_v|+|\S_u|) \cdot D_{\e} \\
& \C_{16} 	= \sum_{\e_{k}=(v,u) \in \E}  \mychoose{D_{\e}}{2} 
\end{align}
\noindent
Recall that all unrestricted counts $\C_{3},...,\C_{16}$ are computed in $\O(\m)$ time and easily parallelized.

\subsection{Global Graphlet Frequencies} \label{sec:global-graphlet-frequencies}
\noindent
Now, using the above unrestricted counts, we can derive the \emph{connected} and \emph{disconnected} global (macro-level) graphlet counts of size $k \in \{2,3,4\}$ for the graph $G$ as:
\begin{align} \nonumber
&X_{1} = \m \\ 				\nonumber		
&X_{2} = \mychoose{\n}{2} - \m \\	 \nonumber
&X_{3} = \nicefrac{1}{3} \cdot \C_{3} \\ \nonumber
&X_{4} = \nicefrac{1}{2} \cdot \C_{4} \\ \nonumber
&X_{5} = \C_{5} \\ \nonumber
&X_{6} = \mychoose{\n}{3} - \big( X_{3} + X_{4} + X_{5} \big) \\ \nonumber
& X_{7} = \nicefrac{1}{6} \cdot \C_{7}  \\  \nonumber
& X_{8} = \C_{8} - \C_{7} \\ 
& X_{9} = \nicefrac{1}{2} \big( \C_{9} - 4X_{8} \big) \nonumber \\ \nonumber 
& X_{10} = \nicefrac{1}{4} \cdot \C_{10} \\ \nonumber 
& X_{11} = \nicefrac{1}{3} (\C_{9} - X_{9}) \\ \nonumber
& X_{12} = \C_{12} - \C_{10} \\ \nonumber
& X_{13} = \nicefrac{1}{3} \cdot \big( \C_{13} - X_{9} \big) \\ \nonumber
& X_{14} =\nicefrac{1}{2} \cdot \big( \C_{14} - 6X_7 - 4X_{8} - 2X_{9} - 4X_{10} - 2X_{12} \big)  \\ \nonumber
& X_{15} = \nicefrac{1}{2} \cdot (\C_{15} - 2 X_{12}) \\ \nonumber
& X_{16} = \C_{16} - 2X_{14} \\  \nonumber
& X_{17} = \mychoose{\n}{4} - \sum X_i \quad \text{for } i=7,\ldots,16  \nonumber
\end{align}

\subsection{Complexity}
\noindent
This section gives the space and time complexity.
Let $T_{\max}$ and $S_{\max}$ denote the maximum number of triangles and stars incident to a selected edge $\e \in \E$.
Our algorithm solves the graphlet decomposition problem for $k$-vertex induced subgraphs in: 
\begin{align} \label{eq:computational-complexity-pgd} \nonumber
\O\Bigr( \m \dmax \bigr( T_{\max} + S_{\max} \bigl) \Bigl)
\end{align}
\noindent
Using $\m$ processors (cores, workers), this reduces to $\O(\Delta (\tri_{\max}+\S_{\max}) )$.
\noindent
For the local graphlet problem, finding all graphlets centered at an edge $e=(v,u)$ in $G$ is solved in $\O\big( \d_u ( |\tri| + |\S_u|) \big)$ time.

Given an arbitrary edge $\e_k \in \E$ in $G$, Alg~\ref{alg:3-graphlets-bsearch} computes $\tri$ and $\S_u$ in $\O(\sum_{w \in \N(u)} \log \d_w)$ time.
This is due to the fact that each vertex $w \in \N(u)$ takes $\O(\log \d_w)$ time to check if $(v,w) \in \E$.
However, Alg~\ref{alg:3-graphlets-hash-neigh-iter} finds $\tri$ and $\S_u$ in $\O(\d_v + \d_u)$ time.
In particular, Alg~\ref{alg:3-graphlets-hash-neigh-iter} first marks the neighbors of $v$ in $\O(\d_v)$ time. 
Now, for each $w \in \N(u)$, we check in $o(1)$ time if $(v,w) \in \E$ (using the fast lookup table), as this implies that $w$ completes a triangle with $(v,u)$.
Thus, taking a total of $\O(\d_v + \d_u)$ time.
Note that in terms of space, Alg~\ref{alg:3-graphlets-bsearch} is more efficient, since Alg~\ref{alg:3-graphlets-hash-neigh-iter} requires an additional $\O(\n)$ space for $\hash$. 
Note that each parallel worker maintains a local hash table $\hash$, and thus too expensive for GPUs that have thousands of workers/cores.
Thus, Alg.~\ref{alg:3-graphlets-bsearch} is used for GPUs since they have limited memory while also having many more cores (workers) than CPUs.
Now, given $\tri$ and $\S_u$ for edge $\e_k$, we compute 4-cliques in $\O(\dmax |\tri|)$ time. 
More precisely, the 4-cliques centered at $\e_k \in E$ are computed in exactly $\O(\sum_{w_i \in \tri} \N(w_i))$ time.
In a similar fashion, 
Similarly, we compute 4-cycles in $\O(\dmax |\S_u|)$ time, and more precisely, $\O(\sum_{w_i \in \tri} \N(w_i))$.
 
{
\setlength{\tabcolsep}{2.0pt}
\begin{table}[t!]
\caption{Results demonstrate the effectiveness of the hybrid parallel graphlet decomposition algorithms.
In particular, we find that by leveraging the unique features and advantages of CPUs and GPUs, one can obtain significant speedups over existing methods that leverage only CPUs or GPUs, but not both.
}
\vspace{1mm}

\label{table:exp-perf-results}
\centering
\small
\scriptsize
\begin{tabularx}{
1.0\linewidth
}{@{} r HH c HHH ll HH HHHHHH Hl Hr Hr Hr HHH 
HH 
H 
H 
HHHHHH 
@{}}
\toprule
& 
&
&
& 
&
&
& 
&
& 
& 
&
\multicolumn{6}{c}{}
&
&
&
\multicolumn{6}{c}{\bf  Speedup (times faster)} 
\\

& 
&
&
& 
&
&
& 
&
& 
& 
&
\multicolumn{6}{c}{}
&
&
&
& &
& {\sc Multi-} &
& &
\\

\textbf{graph}
& $|E|$
& $\Delta$
& $\KK$
& $|E|$
& $|E_{\rm cpu}|$
& $|E_{\rm gpu}|$
& $\Delta$
& $\Delta_{\rm gpu}$ & 
\includegraphics[scale=0.04]{graphlets/3-triangle.eps} &
\includegraphics[scale=0.04]{graphlets/3-path.eps} &
\includegraphics[scale=0.04]{graphlets/4-clique.eps} &
\includegraphics[scale=0.04]{graphlets/chordal-cycle.eps} &
\includegraphics[scale=0.04]{graphlets/tailed-triangle.eps} &
\includegraphics[scale=0.04]{graphlets/4-cycle.eps} &
\includegraphics[scale=0.04]{graphlets/4-star.eps} &
\includegraphics[scale=0.04]{graphlets/4-path.eps} 
& \textsc{CPU} 
& $\alpha$ 
& \textsc{GPU} 
& \textsc{GPU} 
& $\textsc{Multi-GPU}$
& $\textsc{GPU\,\,\,\,}$  

& \textsc{Hybrid}
& \textsc{Hybrid} 
\\ 
\midrule

$\mathsf{socfb}$-$\mathsf{Texas84}$ 
& 1.6M & 6312 & 81 & 1.6M & 63.6k & 1.5M & 6312 & 450 
& 11.2M & 302.2M 
& 70.7M & 376.9M & 1.2B & 215.2M & 664.6M & 3.9B 
&  & 0.031 
&  & 4.65$\mathsf{x}$ 		
& & 21.91$\mathsf{x}$  			
&  & 263.26$\mathsf{x}$ &
\\

$\mathsf{socfb}$-$\mathsf{UF}$ & 
1.5M & 8246 & 83 & 1.5M & 73.3k & 1.4M & 8246 & 370 & 
12.1M & 265.9M & 
98M & 433.1M & 708.9M & 186.4M & 778.3M & 874.1M & 
 & 0.05 & 
 & 1.6$\mathsf{x}$ &
 & 55.65$\mathsf{x}$ &
& 165.63$\mathsf{x}$ & 
\\

$\mathsf{socfb}$-$\mathsf{MIT}$ & 251.2k & 708 & 72 & 251.2k & 25.1k & 226.1k & 708 & 266 
& 2.4M & 32.3M 
& 13.7M & 88.5M & 909.4M & 50.9M & 498.2M & 3.8B 
&  & 0.7
&  & 11.98$\mathsf{x}$
&  & 28.47$\mathsf{x}$ 
&  & 106.14$\mathsf{x}$ & 	
\\

$\mathsf{socfb}$-$\mathsf{Stanford3}$ & 
568.3k & 1172 & 91 & 568.3k & 28.4k & 539.9k & 1172 & 365 & 
5.8M & 93.7M & 
37.1M & 225.7M & 658.8M & 150.9M & 600.8M & 1.8B & 
 & 0.05 & 
 & 21.07$\mathsf{x}$ &
 & 63$\mathsf{x}$ &
& 133.15$\mathsf{x}$ & 
\\

$\mathsf{socfb}$-$\mathsf{Wisc87}$ & 
835.9k & 3484 & 60 & 835.9k & 33.4k & 802.5k & 3484 & 300 & 
4.9M & 106.8M & 
23M & 121M & 1.9B & 59.3M & 1.3B & 3.8B & 
& 0.04 & 
 	& 17.88$\mathsf{x}$ & 
 	& 142.41$\mathsf{x}$ & 
 	& 189.08$\mathsf{x}$ & 
\\

$\mathsf{socfb}$-$\mathsf{Indiana}$ & 
1.3M & 1358 & 76 & 1.3M & 52.2k & 1.3M & 1358 & 329 & 
9.4M & 180.5M & 
60.2M & 269.1M & 1.6B & 140.6M & 494.6M & 3.9B & 
 & 0.04 & 
& 22.25$\mathsf{x}$ &
 & 96.89$\mathsf{x}$ & 
 & 207.11$\mathsf{x}$ & 
\\

\midrule

$\mathsf{soc}$-$\mathsf{flickr}$
& 3.2M & 4369 & 309 & 3.2M & 0 & 3.2M & 4369 & 4196 
& 58.8M & 962.7M 
& 311.2M & 1B & 208.4M & 252.2M & 1.2B & 3.7B 
&  & 0.04
& & 7.32$\mathsf{x}$	
& & 31.85$\mathsf{x}$ 	
&  & 102.24$\mathsf{x}$ & 			
\\

$\mathsf{soc}$-$\mathsf{google}$-$\mathsf{plus}$
& 1.1M & 1790 & 135 & 1.1M & 114.2k & 1M & 1790 & 328 
& 12.2M & 116.6M 
& 185.8M & 993.9M & 203.6M & 463M & 667.5M & 3.7B 
&  & 0.07
& & 4.95$\mathsf{x}$ 
&  & 11.98$\mathsf{x}$ 
&  & 56.03$\mathsf{x}$ &
\\

$\mathsf{soc}$-$\mathsf{youtube}$ & 
1.9M & 25409 & 49 & 1.9M & 135.6k & 1.8M & 25409 & 1079 & 
2.4M & 825.1M & 
3.8M & 155.5M & 1.2B & 162.4M & 1B & 2.3B & 
 & 0.07 & 
 & 3.87$\mathsf{x}$ &
 & 26.82$\mathsf{x}$ &
 & 180.64$\mathsf{x}$ & 
\\

$\mathsf{soc}$-$\mathsf{brightkite}$ & 212.9k & 1134 & 52 & 212.9k & 25.6k & 187.4k & 1134 & 132 
& 494.4k & 11.9M 
& 2.9M & 12.2M & 114M & 2.7M & 1.3B & 532.6M 
&  & 0.12 
& & 2.51$\mathsf{x}$ 
&  & 8.09$\mathsf{x}$ 
&  & 17.67$\mathsf{x}$ &
\\

$\mathsf{soc}$-$\mathsf{livejournal}$ & 
27.9M & 2651 & 213 & 27.9M & 2.5M & 25.4M & 2651 & 157 & 
83.6M & 1.5B & 
307M & 1.9B & 1.8B & 465.3M & 778.1M & 3.5B & 
 & 0.05 & 
& 8.92$\mathsf{x}$ & 
 & 70.01$\mathsf{x}$ & 
 & 98.83$\mathsf{x}$ & 
\\

$\mathsf{soc}$-$\mathsf{twitter}$ & 
12.5M & 51386 & 125 & 12.5M & 625.4k & 11.9M & 51386 & 13533 & 
83M & 620.8M & 
429.7M & 2.3B & 1.7B & 990.1M & 313.5M & 1.9B & 
 & 0.05 & 
 & 2.68$\mathsf{x}$ &
 & 21.76$\mathsf{x}$ &
& 372.72$\mathsf{x}$ & 
\\ 

$\mathsf{soc}$-$\mathsf{orkut}$ & 
106.3M & 27466 & 230 & 106.3M & 8.5M & 97.8M & 27466 & 646 & 
524.6M & 934.3M & 
280.2M & 3.2B & 952.8M & 594.7M & 520.3M & 2B & 
 & 0.05 & 
& 6.12$\mathsf{x}$ &
 & 57.71$\mathsf{x}$ &
 & 129.26$\mathsf{x}$ & 
\\

\midrule

$\mathsf{ia}$-$\mathsf{enron}$-$\mathsf{large}$ & 180.8k & 1383 & 43 & 180.8k & 31.8k & 149k & 1383 & 243 
& 725.3k & 23.4M 
& 2.3M & 22.5M & 375.7M & 6.8M & 184.6M & 1.4B 
&  & 0.176 
& & 2.94$\mathsf{x}$
& & 10.79$\mathsf{x}$
&  & 28.30$\mathsf{x}$ & 
\\

$\mathsf{ia}$-$\mathsf{wiki}$-$\mathsf{Talk}$
& 360.8k & 1220 & 58 & 360.8k & 0 & 360.8k & 1220 & 1034 & 836.5k & 51.1M & 2.2M & 32.3M & 667.8M & 33.8M & 766.2M & 1.5B 
&  & 0.02 
&  & 23.35$\mathsf{x}$
&  & 37.50$\mathsf{x}$
&  & 85.46$\mathsf{x}$ &
\\

\midrule

$\mathsf{ca}$-$\mathsf{HepPh}$ & 117.6k & 491 & 238 & 117.6k & 41.2k & 76.5k & 491 & 169 
& 3.4M & 5.2M 
& 150.3M & 35.2M & 462.3M & 820.7k & 143.2M & 203.8M 
& & 0.35 
& & 1.42$\mathsf{x}$
&  & 6.62$\mathsf{x}$
&  & 17.14$\mathsf{x}$ &
\\

\midrule

$\mathsf{brain}$-$\mathsf{mouse}$-$\mathsf{ret1}$ 
& 90.8k & 744 & 121 & 90.8k & 0 & 90.8k & 744 & 712 & 3.3M & 14.8M & 71.4M & 302.9M & 1.1B & 47.4M & 1.1B & 1.1B 
&  & 0.26 
& & 3.21$\mathsf{x}$
&  & 5.14$\mathsf{x}$ 
&  & 32.71$\mathsf{x}$ &
\\

\midrule

$\mathsf{web}$-$\mathsf{baidu}$-$\mathsf{baike}$ & 
17M & 97848 & 78 & 17M & 510.4k & 16.5M & 97848 & 11919 & 
25.2M & 744.5M & 
27.8M & 247.9M & 476.2M & 653M & 1.3B & 1.2B & 
& 0.03 & 
 &  4.83$\mathsf{x}$ & 
 & 39.55$\mathsf{x}$ & 
 & 156.45$\mathsf{x}$ & 
\\

$\mathsf{web}$-$\mathsf{arabic05}$ & 1.7M & 1102 & 101 & 1.7M & 262.1k & 1.5M & 1102 & 49 
& 21.7M & 3.6M 
& 232.3M & 3.4M & 26.5M & 79.2k & 489.6M & 27.3M 
&  & 0.14 
& & 5.19$\mathsf{x}$
&  & 29.51$\mathsf{x}$
&  & 60.02$\mathsf{x}$ &
\\

\midrule

$\mathsf{tech}$-$\mathsf{internet}$-$\mathsf{as}$ & 85.1k & 3370 & 23 & 85.1k & 29.8k & 55.3k & 3370 & 208 & 63.3k & 28.4M & 84.6k & 5.1M & 105.6M & 1.5M & 796.1M & 642.5M 
&  & 0.35 
& & 1.26$\mathsf{x}$
&  & 3.55$\mathsf{x}$
& & 12.78$\mathsf{x}$ &
\\

$\mathsf{tech}$-$\mathsf{as}$-$\mathsf{skitter}$
& 11.1M & 35455 & 111 & 11.1M & 1.1M & 10M & 35455 & 4768 
& 28.8M & 903M 
& 148.8M & 2.4B & 570.9M & 816.9M & 807.6M & 2.8B 
&  & 0.08 
&  & 0.62$\mathsf{x}$
&  & 1.89$\mathsf{x}$ 
&  & 58.62$\mathsf{x}$ & 
\\

\midrule

$\mathsf{C500}$-$\mathsf{9}$ 
& 112.3k & 468 & 432 & 112.3k & 47.2k & 65.2k & 468 & 450 & 15.1M & 5M & 655.9M & 909.4M & 201.4M & 50.2M & 7.3M & 22.3M 
&  & 0.42 
&  & 3.13$\mathsf{x}$	
&  & 21.99$\mathsf{x}$ 
&  & 33.23$\mathsf{x}$ & 
\\

$\mathsf{p}$-$\mathsf{hat500}$-$\mathsf{1}$ & 
31.6k & 204 & 86 & 31.6k & 0 & 31.6k & 204 & 199 & 
419.1k & 3M & 
1.5M & 18.8M & 87.7M & 19.3M & 76.8M & 199.8M & 
 & 0.1 & 
 & 13.8$\mathsf{x}$ &
 & 28.02$\mathsf{x}$ & 
 & 46.23$\mathsf{x}$ &
\\

$\mathsf{p}$-$\mathsf{hat1000}$-$\mathsf{1}$ & 
122.3k & 408 & 163 & 122.3k & 14.7k & 107.6k & 408 & 323 & 
3.1M & 23.2M & 
20.3M & 264.9M & 1.3B & 281.6M & 1.2B & 3B & 
 & 0.1 &
 & 14.95$\mathsf{x}$ &
 & 67.87$\mathsf{x}$ &
 & 117.3$\mathsf{x}$ & 
\\ 
\bottomrule
\end{tabularx}
\end{table}
}

\medskip\noindent\textbf{Space}: 
Each GPU has a copy of the graph taking $\O(|E|+|V|+1)$ space and a set of edges $\I_i$.
In addition, each GPU has an array of length $P_i \cdot \dmax_i$ where $\dmax_i$ is the maximum degree of any vertex $u$ in the set of edges $\I_i$  
and $P_i$ is the total cores (workers/processing units) of the $i^{\th}$ GPU. 
Thus, $\dmax_i$ is an upper bound on the maximum size any edge neighborhood task would require, see Figure~\ref{fig:graphlets-intuition} for intuition.
Thus, the total space for the $i^{\th}$ GPU is:
\[
\O\Big( |E|+|V| + |\I_i| + (P_i \cdot \dmax_i)  \Big)
\]
\noindent
Note the above is for global (macro) graphlet counts.
Now, suppose we want to compute the local graphlet counts centered at each edge.
This would require three additional arrays all of size $|\I_i|$.
Notice that for global macro counts, each GPU simply communicates the total number of triangles, cliques, and cycles, which can be easily aggregated locally (as opposed to the count of triangles, cliques, and cycles for each edge in $\I_i$).
This avoids expensive and unnecessary communications.

Each CPU worker has a local hash table $\hash$ taking $\O(\n)$ space, as well as two arrays for storing $\tri$ and $\S_u$ of length $\dmax$, thus the total space per CPU worker is $\O(\n + 2\dmax)$.
Assuming $P$ workers, the total space is: $\O(P (\n + 2\dmax)) = \O(P\n)$.
The graph is $\O(|V|+|E|+1)$ and shared among the CPU workers.

{
\setlength{\tabcolsep}{4.0pt}
\begin{table}[b!]
\vspace{-1.5mm}
\caption{\emph{Connected 4-graphlet} frequencies for a variety of the real-world networks investigated from different domains.
}
\vspace{0.8mm}

\label{table:exp-perf-results-stats}
\centering
\scriptsize

\begin{tabularx}{\linewidth}{
@{}
r 
HH H HHH 
HH 
HH 
XXXXXX H 
HH HH HH HHH
HH 
H 
H 
HHHHHH 
@{}}

\toprule
& 
&
&
& 
&
&
& 
&
& 
\multicolumn{8}{c}{\bf  Connected Graphlets} 
\vspace{1.1mm}
&
&
&
\\

\textbf{graph}
& & & & & & & & & 
\includegraphics[scale=0.04]{graphlets/3-triangle.eps} &
\includegraphics[scale=0.04]{graphlets/3-path.eps} &
\includegraphics[scale=0.04]{graphlets/4-clique.eps} &
\includegraphics[scale=0.04]{graphlets/chordal-cycle.eps} &
\includegraphics[scale=0.04]{graphlets/tailed-triangle.eps} &
\includegraphics[scale=0.04]{graphlets/4-cycle.eps} &
\includegraphics[scale=0.04]{graphlets/4-star.eps} &
\includegraphics[scale=0.04]{graphlets/4-path.eps} 
\\ 
\midrule

$\mathsf{socfb}$-$\mathsf{Texas84}$ 
& 1.6M & 6312 & 81 & 1.6M & 63.6k & 1.5M & 6312 & 450 
& 11.2M & 302.2M 
& 70.7M & 376M & 1.2B & 215M & 664M & 3.9B 
\\

$\mathsf{socfb}$-$\mathsf{UF}$ & 
1.5M & 8246 & 83 & 1.5M & 73.3k & 1.4M & 8246 & 370 & 
12.1M & 265.9M & 
98M & 433M & 708M & 186M & 778M & 874M 
\\

$\mathsf{socfb}$-$\mathsf{MIT}$ & 251.2k & 708 & 72 & 251.2k & 25.1k & 226.1k & 708 & 266 
& 2.4M & 32.3M 
& 13.7M & 88.5M & 909M & 50.9M & 498M & 3.8B 
\\

$\mathsf{socfb}$-$\mathsf{Stanford3}$ & 
568.3k & 1172 & 91 & 568.3k & 28.4k & 539.9k & 1172 & 365 & 
5.8M & 93.7M & 
37.1M & 226M & 659M & 151M & 600M & 1.8B 
\\

$\mathsf{socfb}$-$\mathsf{Wisc87}$ & 
835.9k & 3484 & 60 & 835.9k & 33.4k & 802.5k & 3484 & 300 & 
4.9M & 107M & 
23M & 121M & 1.9B & 59.3M & 1.3B & 3.8B 
\\

$\mathsf{socfb}$-$\mathsf{Indiana}$ & 
1.3M & 1358 & 76 & 1.3M & 52.2k & 1.3M & 1358 & 329 & 
9.4M & 181M & 
60.2M & 269M & 1.6B & 141M & 495M & 3.9B 
\\
\midrule

$\mathsf{soc}$-$\mathsf{flickr}$
& 3.2M & 4369 & 309 & 3.2M & 0 & 3.2M & 4369 & 4196 
& 58.8M & 963M 
& 311M & 1B & 208M & 252M & 1.2B & 3.7B 		
\\

$\mathsf{soc}$-$\mathsf{google}$-$\mathsf{plus}$
& 1.1M & 1790 & 135 & 1.1M & 114k & 1M & 1790 & 328 
& 12.2M & 117M 
& 186M & 994M & 204M & 463M & 668M & 3.7B 
\\

$\mathsf{soc}$-$\mathsf{youtube}$ & 
1.9M & 25409 & 49 & 1.9M & 136k & 1.8M & 25409 & 1079 & 
2.4M & 825M & 
3.8M & 156M & 1.2B & 162M & 1B & 2.3B 
\\

$\mathsf{soc}$-$\mathsf{livejournal}$ & 
27.9M & 2651 & 213 & 27.9M & 2.5M & 25.4M & 2651 & 157 & 
83.6M & 1.5B & 
307M & 1.9B & 1.8B & 465M & 778M & 3.5B 
\\

$\mathsf{soc}$-$\mathsf{twitter}$ & 
12.5M & 51386 & 125 & 12.5M & 625.4k & 11.9M & 51386 & 13533 & 
83M & 621M & 
430M & 2.3B & 1.7B & 990M & 314M & 1.9B 
\\

$\mathsf{soc}$-$\mathsf{orkut}$ & 
106.3M & 27466 & 230 & 106M & 8.5M & 97.8M & 27466 & 646 & 
525M & 934M & 
280M & 3.2B & 953M & 595M & 520M & 2B 
\\ 
\midrule

$\mathsf{ia}$-$\mathsf{enron}$-$\mathsf{large}$ & 180.8k & 1383 & 43 & 181k & 31.8k & 149k & 1383 & 243 
& 725k & 23.4M 
& 2.3M & 22.5M & 376M & 6.8M & 185M & 1.4B 
\\

$\mathsf{ia}$-$\mathsf{wiki}$-$\mathsf{Talk}$
& 361k & 1220 & 58 & 361k & 0 & 361k & 1220 & 1034 & 837k & 51.1M & 2.2M & 32.3M & 668M & 33.8M & 766M & 1.5B 
\\
\midrule

$\mathsf{ca}$-$\mathsf{HepPh}$ & 118k & 491 & 238 & 118k & 41.2k & 76.5k & 491 & 169 
& 3.4M & 5.2M 
& 150M & 35.2M & 462M & 821k & 143M & 204M 
\\

\midrule

$\mathsf{brain}$-$\mathsf{mouse}$-$\mathsf{ret1}$ 
& 90.8k & 744 & 121 & 90.8k & 0 & 90.8k & 744 & 712 & 3.3M & 14.8M & 71.4M & 303M & 1.1B & 47.4M & 1.1B & 1.1B 
\\

\midrule

$\mathsf{web}$-$\mathsf{baidu}$-$\mathsf{baike}$ & 
17M & 97848 & 78 & 17M & 510k & 16.5M & 97848 & 11919 & 
25.2M & 745M & 
27.8M & 248M & 476M & 653M & 1.3B & 1.2B & 
\\

$\mathsf{web}$-$\mathsf{arabic05}$ & 1.7M & 1102 & 101 & 1.7M & 262.1k & 1.5M & 1102 & 49 
& 21.7M & 3.6M 
& 232M & 3.4M & 26.5M & 79.2k & 490M & 27.3M 
\\

\midrule

$\mathsf{tech}$-$\mathsf{as}$-$\mathsf{skitter}$
& 11.1M & 35455 & 111 & 11.1M & 1.1M & 10M & 35455 & 4768 
& 28.8M & 903M 
& 149M & 2.4B & 571M & 817M & 808M & 2.8B 
\\

\midrule

$\mathsf{C500}$-$\mathsf{9}$ 
& 112.3k & 468 & 432 & 112.3k & 47.2k & 65.2k & 468 & 450 & 15.1M & 5M & 656M & 909M & 201M & 50.2M & 7.3M & 22.3M 
\\

$\mathsf{p}$-$\mathsf{hat1000}$-$\mathsf{1}$ & 
122.3k & 408 & 163 & 122.3k & 14.7k & 107.6k & 408 & 323 & 
3.1M & 23.2M & 
20.3M & 265M & 1.3B & 282M & 1.2B & 3B 
\\ 

\bottomrule
\end{tabularx}
\end{table}
}

\subsection{Representative Methods from Framework}
\noindent
Observe that our approach succinctly generalizes across the spectrum of GPU graphlet methods, including the following three classes:
\smallskip
\begin{compactitem}
\setlength{\parskip}{4pt}
\item Single GPU algorithm uses only a single GPU device for computing induced subgraphs of size $k\in\{2,3,4\}$.

\item Multi-GPU algorithm that leverages multiple GPU devices for graphlets.

\item Hybrid Multi-core CPU-GPU algorithm that leverages the unique computing capabilities of each type of processing unit.
\end{compactitem}
\medskip
\noindent
For instance, if we manually set the upper bound on the local CPU queue to $\alpha=0$, then this gives rise to the Multi-GPU algorithm that leverages \emph{only} GPU devices.
Similarly, if we manually set $\alpha=0$ \emph{and} set the maximum number of GPU devices to $1$ (\ie, \texttt{--}$\mathtt{GPUs}$ $1$), then we have the single GPU algorithm.
We investigate each of these methods in Section~\ref{sec:exp} and evaluate them against the state-of-the-art CPU parallel graphlet decomposition ($\pgd$) framework~\cite{pgd-kais,nkahmed:icdm15}.

\section{Experiments} \label{sec:exp}
\noindent
This experiments in this section are designed to answer the following two questions that lie at the heart of this work.
First, does the GPU and multi-GPU only algorithms improve performance over the state-of-the-art CPU method?
Second, can we leverage the unique features and advantages of both CPUs and GPUs to further improve performance? 

For these experiments, two Intel Xeon CPU E5-2687 @ 3.10GHz were used with 8 cores each.
Further, we used 8 GeForce GTX TITAN Black GPUs and each GPU has 2880 cores (889 MHz base) and 6144 MB memory.
We demonstrate the effectiveness of our approach on a variety of real-world networks from a range of domains with different properties.
We evaluate three parallel graphlet methods from the proposed framework including:
Single GPU only, 
Multi-GPUs, and 
Hybrid approach that uses multiple CPUs and GPUs.
Table~\ref{table:exp-perf-results} shows the speedup relative to the state-of-the-art method called $\pgd$~\cite{pgd-kais,nkahmed:icdm15}.
These results demonstrate the effectiveness of the proposed methods.
In particular, the multi-GPU only and hybrid (CPU+GPU) are orders of magnitude faster than $\pgd$.
For certain types of real-world networks, we find that the methods are over 100 times faster, with the largest improvement being 372$\mathsf{x}$ for $\mathsf{soc}$-$\mathsf{twitter}$. 
In Table~\ref{table:exp-perf-results}, $\dmax_{\rm gpu}$ represents the maximum degree of an edge assigned to the GPUs.
Strikingly, we observe that $\dmax_{\rm gpu}$ is usually much less than $\dmax$.
Graphlet statistics for a few of the graphs are shown in Table~\ref{table:exp-perf-results-stats}.

\begin{figure}[t!]
\centering
\includegraphics[width=0.9\linewidth]{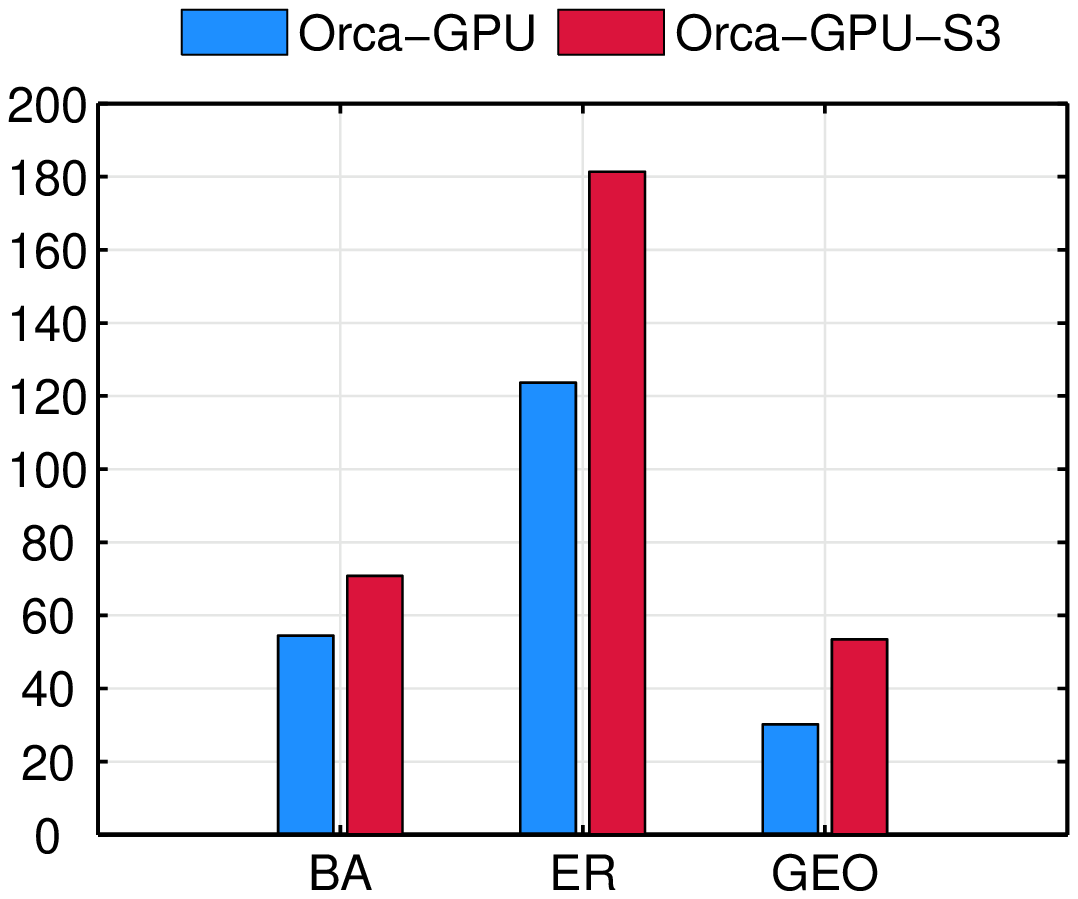}
\caption{
Runtimes of the existing single-GPU method is normalized \wrt the runtime of our approach. 
For the BA graph, our approach is $\approx$55x faster than Orca-GPU, and $\approx$71x faster than Orca-GPU-S3.
Furthermore, for the ER graph, our approach is $\approx$124x faster than Orca-GPU, and $\approx$181x faster than Orca-GPU-S3.
}
\label{fig:Orca-GPU}
\end{figure}

In addition, we also compare to the approach proposed by \citet{milinkovic14contribution}, see Figure~\ref{fig:Orca-GPU}.
Recall that approach is not hybrid (as the one proposed in this work) nor does it use multiple GPUs.
Furthermore, we compute all connected \emph{and} disconnected graphlets up to size $k=4$ (thus including $k\in\{2,3,4\}$), whereas \cite{milinkovic14contribution} is only able to compute \emph{connected graphlets} of size $k=4$ --- a much smaller subset.
Despite this difference, our approach is still orders of magnitude faster as shown in Figure~\ref{fig:Orca-GPU}.
For comparison, we report results using the same benchmark graphs (see \cite{milinkovic14contribution}). 
In particular, that work used random graphs generated from Barab\'{a}si-Albert (BA), Erd\H{o}s-R\'{e}nyi (ER) and geometric algorithms (GEO).
From each, we selected the largest graph for comparison (1K vertices and 150K edges).
Strikingly, our approach is orders of magnitude faster with up to 181x improvement.
It is important to note that our approach is significantly faster for real-world networks where the work associated with each edge is fundamentally unbalanced and highly skewed.
The random graphs in Figure~\ref{fig:Orca-GPU} represent prime candidates for that approach.
Observe that unlike many real-world networks, the degree distribution of these synthetic graphs is not skewed (does not obey a power-law) and are relatively dense ($30\%$).
Nevertheless, our approach is still orders of magnitude faster.

\newcommand{\vol}{\ensuremath{\mathtt{vol}}}

{
\setlength{\tabcolsep}{8.0pt}
\begin{table}[h!]
\caption{
Varying edge ordering can significantly impact performance.
Results demonstrate the effectiveness of the initial ordering technique.}
\vspace{1mm}
\label{table:exp-varying-ordering}
\centering
\small
\begin{tabularx}{
0.96\linewidth}{r H cHcH HH cHcH HH}
\toprule
& 
\multicolumn{4}{c}{\sc Descending} 
& 
\multicolumn{7}{l}{\sc Reverse} 
\\

\textbf{graph}
&
\TT \BB 
& $\vd$
& $\alpha$
& $\vol$
& $\alpha$
& ${\bf rand}$
& $\alpha$
& $\vd^{-1}$
& $\alpha$
& $\vol^{-1}$
& $\alpha$
\\ 
\midrule

$\mathsf{socfb}$-$\mathsf{Texas84}$ & &
263.3$\mathsf{x}$  &  0.031 &
\textbf{284.1}$\mathsf{x}$  &  0.028 & 
18.7$\mathsf{x}$  & 0.028 &
23.5$\mathsf{x}$  & 0.031 & 
10.8$\mathsf{x}$  & 0.028 & 
\\

\bottomrule
\end{tabularx}
\end{table}
}

Table~\ref{table:exp-varying-ordering} demonstrates the impact of various edge orderings (using the Hybrid GPU-CPU approach). 
In particular, we investigate ordering edges from \emph{largest to smallest} degree $\vd$ and volume $\vol$ (\ie, $\vol(\e_k)=\vol(u,v)=\sum_{w\in \N(u,v)} \d_w$, which is the sum of degrees of vertices in $\N(u,v)$).
The impact of the reverse ordering is also investigated, $\ie$, ordering edges from smallest to largest degree and degree volume denoted by $\vd^{-1}$ and $\vol^{-1}$, respectively.
Notably, we observe that the ordering can significantly influence performance and in some instances may even lead to slower performance than a GPU-only approach.

\begin{figure}[h!]
\centering
\includegraphics[width=0.95\linewidth]{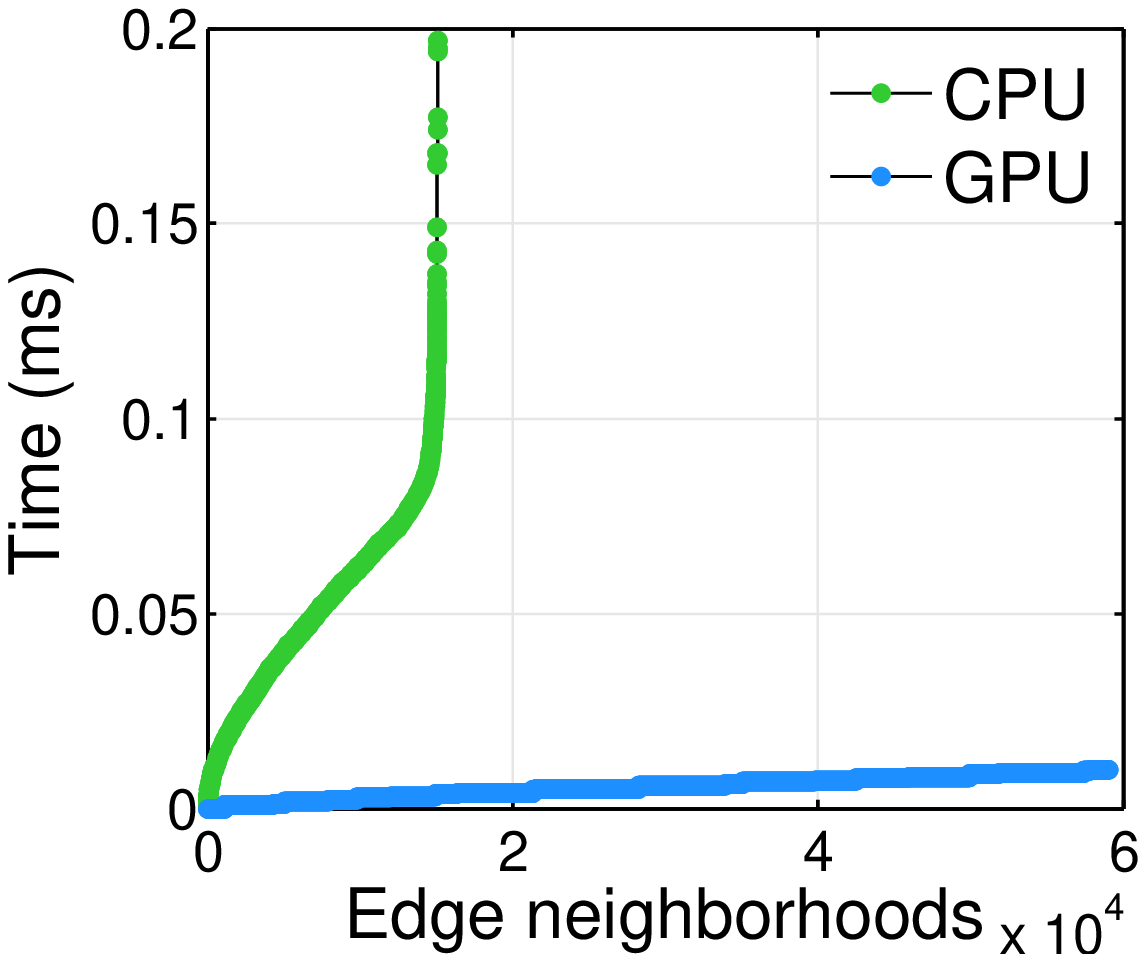}
\caption{
CPU and GPU processing time for each edge neighborhood.
The x-axis represents edge neighborhoods which are computed by either a CPU or GPU, whereas the y-axis is the time (ms) for computing each edge neighborhood. 
See text for discussion.
}
\label{fig:graphlet-cpu-gpu-runtime}
\end{figure}

Figure~\ref{fig:graphlet-cpu-gpu-runtime} validates the proposed hybrid (GPU+CPU) approach.
Recall that our framework dynamically partitions the edges among the CPUs and GPUs based on some notion of difficulty.
In particular, we see that edge neighborhoods assigned to the CPU are indeed difficult and require significantly more time to compute than the edge neighborhoods processed by the GPUs.
In addition, our approach is more space-efficient for the GPU, which has significantly less RAM than the CPU.
Recall that GPUs are assigned neighborhoods that are significantly more sparse than those given to the CPUs, and therefore requires less space as well as avoiding expensive communications.
Results in Figure~\ref{fig:graphlet-cpu-gpu-runtime} also demonstrate the effectiveness of the dynamic load balancing approach used in the hybrid graphlet algorithm, as it assigns edge neighborhoods to the corresponding ``best" processor type. 
Moreover, the above also demonstrates the effectiveness and importance of the initial edge ordering $\Pi$. 
Notably, we found degree to be a useful approximation of the actual work required to compute the local graphlet counts for each edge neighborhood.
However, ordering edges by $\vol$ can lead to improvements in performance over degree (Table~\ref{table:exp-varying-ordering}).
Moreover, we also exploit different graphlet algorithms (that essentially trade-off space for time) by dynamically selecting the appropriate one at runtime based on simple heuristics that can be derived in constant time. 
The GPUs are then used to compute graphlets for edge neighborhoods that are more well-balanced and regular, which is exactly the type of problems for which they are most effective.

Finally, memory use for GPUs is shown in Figure~\ref{fig:graphlet-cpu-gpu-memory-mb}. 
Notice that the graph usually exceeds the others.
For as-skitter, the memory used to store $\tri$ and $\S_u$ for each GPU worker is slightly larger due to the large degrees.
Finally, as expected the set of edges $\I_i$ is always less than the others.
We also note that the communication overhead is insignificant compared to the time to compute graphlets.

\begin{figure}[t!]
\centering
\includegraphics[width=0.75\linewidth]{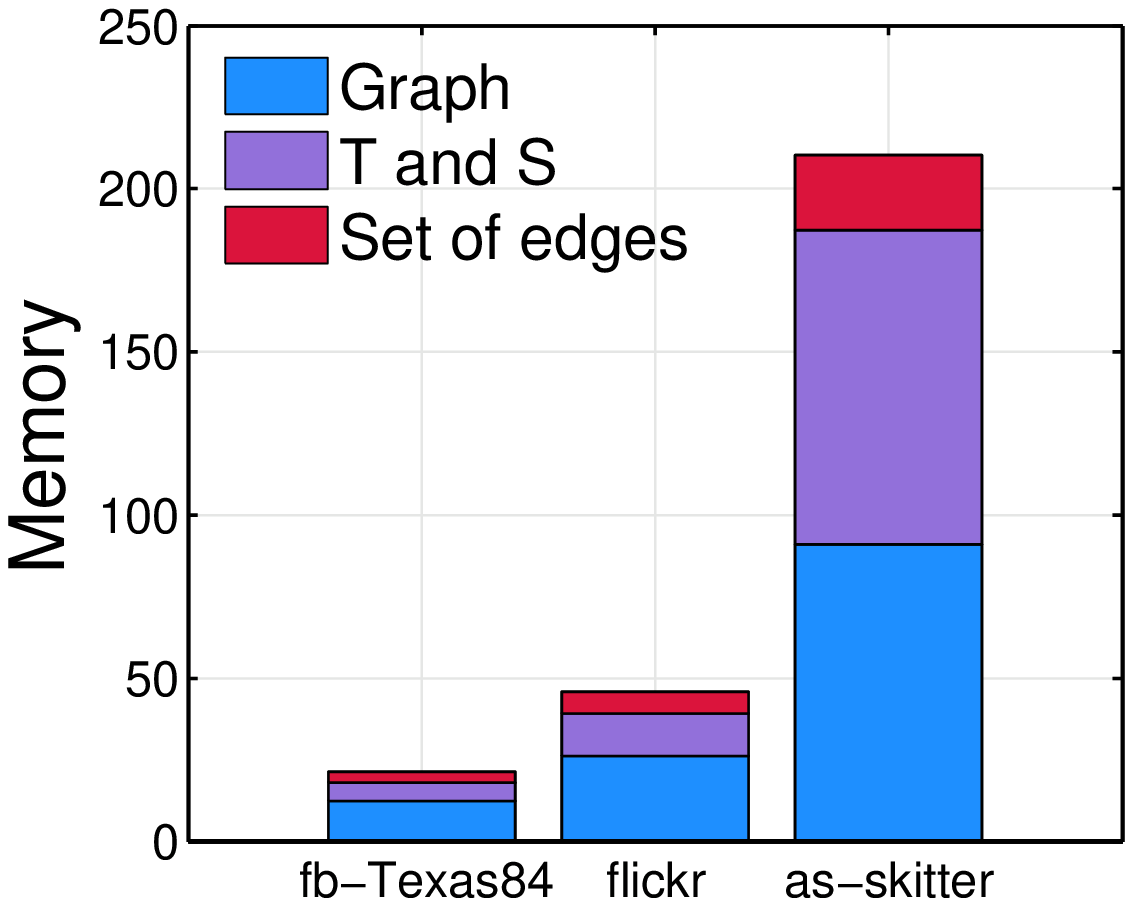}
\vspace{-2mm}
\caption{Average memory (MB) per GPU for three real-world networks.}
\label{fig:graphlet-cpu-gpu-memory-mb}
\vspace{-2mm}
\end{figure}

\section{Conclusion} \label{sec:conc}
\noindent
This work proposed a parallel graphlet decomposition method that effectively leverages multiple CPUs and GPUs simultaneously.
The algorithm is designed to exploit the unique features and strengths of each type of processing unit and is shown to be orders of magnitude faster than existing work that is based on only a single type of computing device. 
In particular, the proposed methods were shown to be up to 300+ times faster than the recent state-of-the-art.
Besides being orders of magnitude faster, our approach is also more energy efficient.
The proposed methods are also well-suited for unbiased graphlet estimation, and we plan to investigate this problem in future work.

\bibliographystyle{abbrvnat}
\bibliography{paper}
\end{document}